\documentclass[a4paper,11pt]{article}
\usepackage[colorlinks=true,linkcolor=blue, citecolor=blue, urlcolor=blue]{hyperref}
\usepackage{amsmath}
\usepackage{amssymb}
\usepackage{amsfonts}
\usepackage[pdftex]{graphicx}
\usepackage{color}
\usepackage{lscape}
\usepackage{bm}
\usepackage{comment}
\usepackage{natbib}
\usepackage{adjustbox}
\newtheorem{theorem}{Theorem}
\newtheorem{lemma}[theorem]{Lemma}

\newtheorem{example}{\it Example}
\newtheorem{proposition}[theorem]{Proposition}

\newcommand{\dis}{\displaystyle}

\title{An extension of sine-skewed circular distributions}
\author{Yoichi Miyata\thanks{Faculty of Economics, Takasaki City University of Economics, 1300 Kaminamie, Takasaki, Gunma 370-0801, Japan}\and Takayuki Shiohama\thanks{Department of Data Science, Nanzan University, 18 Yamazato-cho, Showa, Nagoya 466-8673, Japan}\and Toshihiro Abe\thanks{Faculty of Economics, Hosei University, 4342 Aihara, Machida, Tokyo, 194-0298, Japan}}

\begin{document}
\maketitle

\hspace*{-1.5em}{\bf Key words}: Asymmetry; Circular statistics; Sine-skewed circular distributions

\begin{abstract}
Sine-skewed circular distributions are identifiable and have easily-computable trigonometric moments and a simple random number generation algorithm, whereas they are known to have relatively low levels of asymmetry. This study proposes a new family of circular distributions that can be skewed more significantly than that of existing models. It is shown that a subfamily of the proposed distributions is identifiable with respect to parameters and all distributions in the subfamily have explicit trigonometric moments and a simple random number generation algorithm. The maximum likelihood estimation for model parameters is considered and its finite sample performances are investigated by numerical simulations. Some real data applications are illustrated for practical purposes.
\end{abstract}

\section{Introduction}\label{Intro}
Owing to the need to predict and fit asymmetric patterns of real datasets, increasing attention has been paid to analyzing asymmetric circular data in recent years.  Examples of early studies on asymmetric circular distributions include \cite{batschelet1981circular} and \cite{yfantis1982extension}. In the last two decades, many researchers, such as \\
\citet{UJ09}, \citet{AP11}, \citet{LV17a}, \citet{JK04}, \citet{KS13} and \citet{KJ10}, have proposed and investigated asymmetric distributions on the circle. 
\citet{LBC21} list five properties that a flexible statistical model should have: 1. Versatility, 2. Tractability, 3. Interpretability, 4. Data generating mechanism, and 5. Straightforward parameter estimation.
Tractability refers to the fact that the density function does not contain complex terms such as infinite sums, and is in a form that is easy to compute. Versatility refers to the ability to give the density function as many different shapes as possible. Interpretability means that parameters should have clear interpretations to infer conclusions about the underlying population. Therefore, it is natural to consider the identifiability to be included in this interpretability because in a nonidentifiable family, two different parameters represent the same probability distribution, which makes the interpretation of the parameters more difficult.

The sine skewed circular distributions proposed by \citet{AP11} are skew-symmetric distributions on the circle satisfying the above properties 1-5. In particular, since these distributions have very simple normalizing constants, they are tractable in the sense of 2. On the other hand, it is difficult for them to be largely asymmetric near the modes of the distributions. When the sine-skewed circular distributions are applied to such asymmetric data, the value of the maximum likelihood estimator for the skewness parameter sometimes occurs at the boundary of the parameter space. In such a case, the estimated probability model may not adequately capture the asymmetry of the data itself, and the asymptotic variance of the estimator may not be properly evaluated.

To address this problem, \citet{BRAL22} proposed a generalized circular sine-skewed distribution by adopting the idea of \citet{Ba02} to derive a generalized skew-normal distribution. This distribution has a skewing function in the form of a polynomial of the sine function and can give a wider range of skewness than the sine skewed circular distributions. In addition, the normalizing constant is expressed in an explicit form, and random numbers are relatively easily generated from this distribution. On the other hand, the normalizing constant depends on base densities and some parameters, and the degree of skewness that this distribution can attain is not strong enough. To overcome these shortcomings, this paper presents another extension of the sine-skewed circular distributions.

%Therefore, in this study, we propose a new model that extends the sine-skewed circular distributions using the method given by \citet{LV17a} and \citet{UJ09}.

Before moving on to the next section, we introduce an approach to construct a generalized asymmetric distribution on the hypersphere ${\cal S}^{p}$ proposed by \citet{LV17a}. Suppose that $\psi (\mu )=(\cos \mu ,\sin \mu )^{T}$, $G :\mathbb{R}\to [0,1]$ is a monotonically increasing continuous function with $G (-y)+G(y)=1$ for any $y\in\mathbb{R}$, and $h_{\bm{\rho}}:[-1,1]\to\mathbb{R}^{+}$ is an absolutely continuous function that is allowed to have an unknown parameter vector $\bm{\rho}$.
Then, $c_{\bm{\rho}}h_{\bm{\rho}}(\psi (\theta )^{T}\psi (\mu ))$ is the symmetric density function of a circular random variable $\Theta$, where $c_{\bm{\rho}}$ is the normalizing constant.
 \citet{LV17a} proposed the following skew-rotationally-symmetric (SRS) distribution with density
\begin{align}
f_{\textrm{SRS}}(\theta ;\mu ,\bm{\rho},\lambda )=2c_{\bm{\rho}}h_{\bm{\rho}}(\psi (\theta )^{T}\psi (\mu ))G \left\{ \lambda \psi (\mu )^{T}\bm{Q} \psi (\theta )\right\} , \label{SRS0}
\end{align}
where $\bm{Q}=\begin{pmatrix}0 & 1\\ -1 & 0\end{pmatrix}$, $\bm{A}^T$ indicates the transpose of matrix $\bm{A}$, and $G\{ \cdot \}$ means the function $G(y)$ with $y$ replaced by $\lambda \psi (\mu )^{T}\bm{Q} \psi (\theta )$. 

For example, if $G (y)=(1+y)/2$ $(y\in [-1,1])$ and $h_{\rho}(y)=1/(1+\rho^{2}-2\rho y)$, it follows that 
\begin{align*}
f_{\textrm{SRS}}(\theta ;\mu ,\bm{\rho},\lambda )&=2c_{\bm{\rho}}h_{\bm{\rho}}(\psi (\theta )^{T}\psi (\mu ))G \left\{ \lambda \psi (\mu )^{T}\bm{Q}\psi (\theta )\right\}  \\
&=2c_{\bm{\rho}}h_{\bm{\rho}}(\psi (\theta )^{T}\psi (\mu ))\left\{ \frac{1+\lambda \psi (\mu )^{T}\bm{Q}\psi (\theta )}{2}  \right\} \\
&=\frac{1 - \rho^2}{2\pi (1 + \rho^2 - 2 \rho \cos (\theta  - \mu ))}\left\{ 1+\lambda \sin (\theta -\mu )\right\},
\end{align*}
which indicates the circular density of the sine-skewed wrapped Cauchy distribution.

This study intends to propose a successful and easily-computable choice of $G$ and $h_{\bm{\rho}}$ such that density \eqref{SRS0} is easily-computable, has cosine and sine moments expressed in explicit forms, and can be skewed more strongly than the sine-skewed circular distributions around the mode.

The remainder of this paper is organized as follows. In Section \ref{Sec2}, we propose skew-symmetric circular distributions that satisfy the conditions of \cite{LV17a}. In Section \ref{Sec3}, we present their cosine and sine moments in explicit forms and show that the identifiability of the family is clarified. We also explain how to generate random numbers from the distributions.
Section \ref{Sec4} gives asymptotic properties of the maximum likelihood estimator, and Section \ref{Sims} describes how to choose a hyper-parameter that needs to be determined a priori to perform the estimation and verifies their performance through some simulations. 
Section \ref{Sec6} presents two real data examples, and Section \ref{Sec7} discusses further applications of the proposed model and issues related to the proposed model.

\section{Extended sine-skewed circular distributions}\label{Sec2}
We now assign a specific function to the function $G$ in the density \eqref{SRS0}. First, we consider a density function of $x$ on the interval $[-1,1]$:
\begin{equation*}
g_{m}(x)=\frac{\Gamma (2(m +1))}{2^{2m +1}\Gamma (m +1)^{2}}(1-x^{2})^{m} \qquad (-1\leq x\leq 1),
\end{equation*}
where $m \geq 0$ is a prespecified value. This density function is a transformation of the density function of beta distribution. Let $\dis G_{m}(x) =\int_{-1}^{x}g_{m}(t)dt$ be its distribution function. 
Then, we propose an extended sine-skewed (ESS) circular distribution of order $m$ with density
\begin{equation}
f_{\textrm{ESS}}^{(m)}(\theta ;\mu ,\bm{\rho},\lambda )=2f_{0}(\theta -\mu ;\bm{\rho})G_{m}\left( \lambda \sin (\theta -\mu )\right) , \label{ESScircular}
\end{equation}
where $f_{0}(\theta ;\bm{\rho})$ is a symmetric base density about zero, and the parameter $m \geq 0$ is referred to as an ``order."
The parameter $\mu \in [-\pi, \pi )$ is a location parameter, $\lambda \in [-1,1]$ is a skewness parameter, and the parameter vector $\bm{\rho}$ plays a role in the other shape such as concentration. 

%Note that the density (\ref{ESScircular}) is a special case of that of \citet{LV17a}. 

The ESS circular distributions are defined for any $m \geq 0$. In particular, it is tractable in the case where $m$ is a nonnegative integer since, for $m =0,1,2,\ldots$, the distribution function $G_m(\cdot)$ is expressed as 
\begin{align}
G_{m}(x)&=\int_{-1}^{x}g_{m}(t)dt \notag\\
 	&=C_{m}\sum_{\ell =0}^{m}\binom{m}{\ell}\frac{1}{2\ell +1}(-1)^{\ell}\left( x^{2\ell +1}+1\right) \notag \\ % \label{Gm1} \notag \\
&=C_{m}\sum_{\ell =0}^{m}\binom{m}{\ell}\frac{1}{2\ell +1}(-1)^{\ell}x^{2\ell +1}+\frac{1}{2}, \label{Gm2}
\end{align}
where  $C_{m}=\dfrac{\Gamma\left\{ 2(m+1)\right\}}{2^{2m+1}\Gamma (m+1)^{2}}$.
%The last equation is derived from $G_{m}(1)=1$. 

\begin{figure}[htb]
\begin{minipage}{.48\linewidth}
\centering
\includegraphics[width=.95\linewidth]{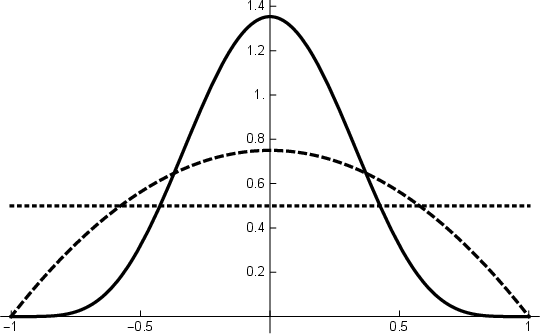}
\caption{Plots of $g_{m}(x)$ $(m =0,1,5)$. ($\cdots$:$m= 0$,- - -:$m=1$, --- :$m=5$)}
\label{fig1}
\end{minipage}
\begin{minipage}{.48\linewidth}
\centering
\includegraphics[width=.95\linewidth]{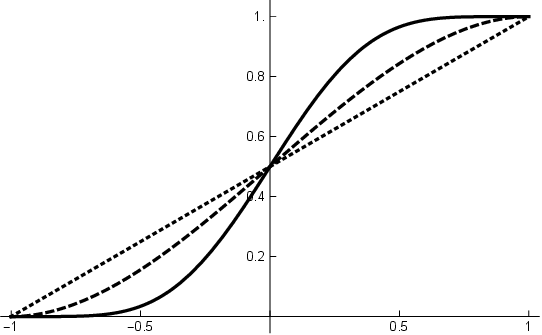}
\caption{Plots of $G_{m}(x)$ $(m =0,1,5)$. ($\cdots$:$m= 0$,- - -:$m=1$, --- :$m=5$)}
\label{fig2}
\end{minipage}
\end{figure}

For example, each of $G_{m}(x)$ $(m =0,1,\ldots ,5)$ is given by
\begin{align*}
G_{0}(x)&=\frac{x+1}{2}, \\
G_{1}(x)&=\frac{1}{4}(-x^{3}+3x+2), \\
G_{2}(x)&=\frac{1}{16}(3x^{5}-10x^{3}+15x+8),  \\
G_{3}(x)&=\frac{1}{32}(-5x^{7}+21x^{5}-35x^{3}+35x+16), \\
G_{4}(x)&=\frac{1}{256}(35 x^9-180 x^7+378 x^5-420 x^3+315 x+128), \quad \textrm{and}\\
G_{5}(x)&=-\frac{1}{512} (x+1)^6 \left(63 x^5-378 x^4+938 x^3-1218 x^2+843 x-256\right).
\end{align*}
%($m>5$\UTF{0088}\UTF{00C8}\UTF{008F}\UTF{00E3}\UTF{0082}\UTF{00C9}\UTF{008A}\UTF{00D6}\UTF{0082}\UTF{00B7}\UTF{0082}\UTF{00E9}\UTF{0083}R\UTF{0083}\UTF{0081}\UTF{0083}\UTF{0093}\UTF{0083}g\UTF{0082}\UTF{00F0}\UTF{0082}±\UTF{0082}±\UTF{0082}\UTF{00C9}\UTF{0093}\UTF{00FC}\UTF{0082}\UTF{00EA}\UTF{0082}\UTF{00E9})}
Recall that the function $G_0(x)$ is the uniform distribution function on $[-1,1]$, which reduces to the sine-skewed circular distribution. If we use a function $G$ with a lower slope around 0, the degree for the skewness parameter $\lambda$ to skew the base density is not strong. 
Hence, we present the skewing function $G_{m}$ having a larger slope around 0 on the support $[-1,1]$. Figures~\ref{fig1} and \ref{fig2} show the functions of $g_m(x)$ and $G_m(x)$, respectively, with different orders $m \in \{0,1,5\}$. 

The density function of the ESS von Mises (ESS-vM) distribution becomes
\begin{equation}
f_{\textrm{ESSvM}}^{(m)}(\theta ;\mu ,\kappa ,\lambda )=\frac{1}{\pi I_{0}(\kappa )}\exp\left\{ \kappa \cos (\theta -\mu )\right\}G_{m}(\lambda \sin (\theta -\mu)), \label{ESSvM}
\end{equation}
where $I_{\nu}(\kappa ):=(1/(2\pi))\int_{0}^{2\pi}\cos (\nu \theta )\exp (\kappa\cos \theta ) d\theta$ is the modified Bessel function of the first kind of order $\nu \in \mathbb{Z}$, $\mu \in [-\pi , \pi )$ is a location parameter, $\kappa > 0$ is a concentration parameter, and $\lambda \in [-1,1]$ is the skewness parameter.
\begin{figure}[htb]
\begin{minipage}{.32\linewidth}
\centering
\includegraphics[width=.99\linewidth, height=4cm]{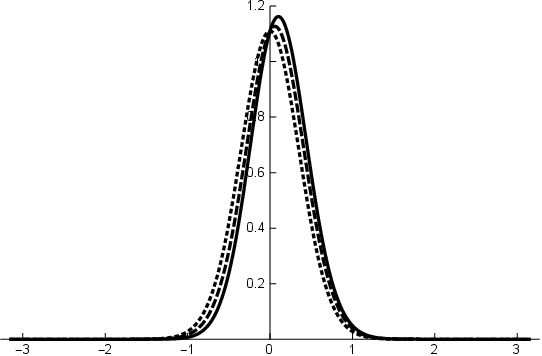}
($m=0$)
\end{minipage}
\begin{minipage}{.32\linewidth}
\centering
\includegraphics[width=.99\linewidth, height=4cm]{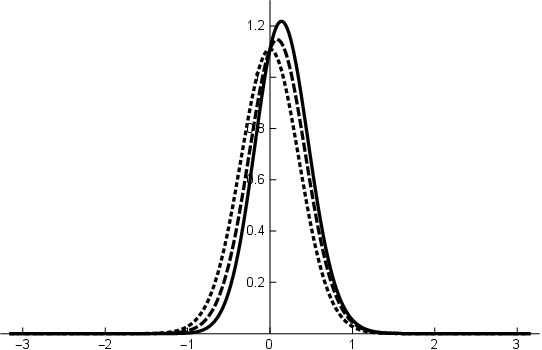}
($m=1$)
\end{minipage}
\begin{minipage}{.32\linewidth}
\centering
\includegraphics[width=.99\linewidth, height=4cm]{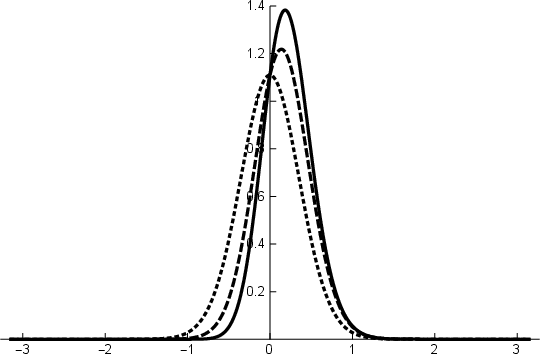}
($m=5$)
\end{minipage}
\caption{Plots of the density function of ESS-vM distributions with parameters $\mu =0$, $\kappa =8$ and different $\lambda$ and $m$. ($\cdots$:$\lambda= 0$,- - -:$\lambda =0.5$, --- :$\lambda=0.9$)}
\label{fig3}
\end{figure}
Figure~\ref{fig3} plots the density function of ESS-vM distributions with parameters $\mu =0$, $\kappa =8$ and different $\lambda$ and $m$. These figures show that as the order $m$ increases, the degree of skewness around the mode also increases. Although \citet[p.691]{AP11} pointed out that the ESS-vM distribution of order $0$ is not unimodal in general, we observed that the ESS-vM distributions of any other order such as $m=1$ and $m=2$ are not also always unimodal.

\section{Some theoretical results}\label{Sec3}
\subsection{Cosine and sine moments}\label{Sec3.1}
Here, we present the $p$th cosine and sine moments of the proposed density $f_{\textrm{ESS}}^{(m)}(\theta ;\mu ,\bm{\rho},\lambda )$. 
To do that, we first consider the case with $\mu =0$. 
Because $G_{m}(\lambda \sin \theta )-1/2$ is an odd function of $\theta$, the $p$th cosine moments under $f_{\textrm{ESS}}^{(m)}(\theta ;0,\bm{\rho},\lambda )$ with $\mu =0$ are 
\begin{align}
\alpha_{p}:=E_{\mu =0}\left\{ \cos (p\Theta )\right\}=\alpha_{0,p} \label{alpha_p},
\end{align}
where $\alpha_{0,p}$ is the $p$th cosine moment under the base density $f_{0}(\theta ;\rho )$ with $\mu =0$. Accordingly, $\alpha_{p}$ does not depend on the order $m$. 
In contrast, from Equation \eqref{Gm2}, the $p$th sine moments under $f_{\textrm{ESS}}^{(m)}(\theta ;0,\bm{\rho},\lambda )$ are expressed as
\begin{align}
\beta_{p}:=2C_{m}\sum_{\ell=0}^{m}\binom{m}{\ell}\frac{(-1)^{\ell}}{2\ell+1}\int_{-\pi}^{\pi}(\lambda \sin \theta )^{2\ell+1}\sin (p\theta )f_{0}(\theta ;\rho )d\theta ,\label{beta_p}
\end{align}
where $f_{0}(\theta ;\rho )$ is the base density with $\mu =0$, and $\binom{0}{0}\equiv 1$ for convenience. 
Then, we define the $p$th mean direction (MD) and $p$th mean resultant length (MRL) by $\mu_{\textrm{MD}}^{(p,m)}:=\textrm{arg}\{ \alpha_{p}+i\beta_{p}\}$ and $\rho_{\textrm{MRL}}^{(p,m)}=\sqrt{\alpha_{p}^2+\beta_{p}^2}$, respectively.

We now describe how to evaluate the integrals in Equation \eqref{beta_p}.
By using 
\begin{align*}
&(\sin \theta )^{3}\sin (p\theta )=\sin (p\theta )\frac{1}{4}\left( 3\sin \theta -\sin (3\theta )\right) \\
&=\frac{3}{8}\left\{ \cos ((p-1)\theta )-\cos((p+1)\theta )\right\} -\frac{1}{8}\left\{  \cos ((p-3)\theta )-\cos ((p+3)\theta )\right\} ,
\end{align*}
the integral with $\ell=1$ is
\begin{align}
\int_{-\pi}^{\pi}(\lambda \sin \theta )^{3}\sin (p\theta )f_{0}(\theta ;\rho )d\theta &=\frac{3\lambda^{3}}{8}\left\{ \alpha_{0,p-1}-\alpha_{0,p+1}\right\} -\frac{\lambda^{3}}{8}\left\{  \alpha_{0,p-3}-\alpha_{0,p+3}\right\} .\label{integral1}
\end{align}
By the same argument as in the above, the integral with $\ell=2$ is given by 
\begin{align}
\int_{-\pi}^{\pi}(\lambda \sin \theta )^{5}\sin (p\theta )f_{0}(\theta ;\rho )d\theta &=\frac{5\lambda^{5}}{16}\left\{ \alpha_{0,p-1}-\alpha_{0,p+1}\right\} -\frac{5\lambda^{5}}{32}\left\{  \alpha_{0,p-3}-\alpha_{0,p+3}\right\} \notag \\
&+\frac{\lambda^{5}}{32}\left\{  \alpha_{0,p-5}-\alpha_{0,p+5}\right\} .\label{integral2}
\end{align}
From these results, if the order is $m=1$, the $p$th cosine and sine moments under $f_{\textrm{ESS}}^{(1)}(\theta ;0,\bm{\rho},\lambda )$ are $\alpha_{p}=\alpha_{0,p}$ and 
\begin{align*}
\beta_{p}=\frac{3}{16}(\lambda^{3}-4\lambda )(\alpha_{0,p+1}-\alpha_{0,p-1})-\frac{\lambda^{3}}{16}(\alpha_{0,p+3}-\alpha_{0,p-3}), \quad (p\in\mathbb{Z}).
\end{align*}
The $p$th MD and $p$th MRL are given by 
\begin{align}
\mu_{\textrm{MD}}^{(p,1)}&=\textrm{arg}\left\{ \alpha_{0,p}+i\left( \frac{3}{16}(\lambda^{3}-4\lambda )(\alpha_{0,p+1}-\alpha_{0,p-1})-\frac{\lambda^{3}}{16}(\alpha_{0,p+3}-\alpha_{0,p-3})\right)\right\} \notag 
\intertext{and}
\rho_{\textrm{MRL}}^{(p,1)}&=\sqrt{\alpha_{0,p}^2+\frac{\lambda^2}{256}\left( 3(\lambda^{2}-4 )(\alpha_{0,p+1}-\alpha_{0,p-1})-\lambda^{2}(\alpha_{0,p+3}-\alpha_{0,p-3})\right)^2} .\notag
\end{align}

If $m=2$, the $p$th cosine and sine moments under $f_{\textrm{ESS}}^{(2)}(\theta ;0,\bm{\rho},\lambda )$ are given by $\alpha_{p}=\alpha_{0,p}$ and 
\begin{align*}
\beta_{p}&=\frac{15}{128}c_{2,1}(\lambda )(\alpha_{0,p+1}-\alpha_{0,p-1}) \\
	&+\frac{5}{256}c_{2,2}(\lambda )(\alpha_{0,p+3}-\alpha_{0,p-3})-\frac{3}{256}\lambda^{5}(\alpha_{0,p+5}-\alpha_{0,p-5}),
\end{align*}
where $c_{2,1}(\lambda )=-\lambda^{5}+4\lambda^{3}-8\lambda$ and $c_{2,2}(\lambda )=3\lambda^{5}-8\lambda^{3}$.

For non-zero location $\mu$, the $p$th cosine and sine moments of $f_{\textrm{ESS}}^{(m)}(\theta ;\mu ,\bm{\rho},\lambda )$ are given by 
\begin{align*}
\alpha_{p,\mu}&=E\left\{ \cos (p\Theta )\right\} 
              =\cos (p\mu )\alpha_{p}-\sin (p\mu )\beta_{p},\qquad \text{and}\\
\beta_{p,\mu}&=E\left\{ \sin (p\Theta )\right\} 
             =\cos (p\mu )\beta_{p}+\sin (p\mu )\alpha_{p}.
\end{align*}
The derivation of these equations is provided in \cite{MSA19}, hence details are omitted. 

We present two examples of the cosine and sine moments for the ESS-vM and ESS wrapped Cauchy (ESS-WC) distributions of order 1. 
\begin{example}\label{ex:moments1}
The ESS-vM distribution of order $1$ with $\mu =0$ has the cosine and sine moments as follows:
\begin{align*}
\alpha_{p}&=\frac{I_{p}(\kappa )}{I_{0}(\kappa )}, \qquad \textrm{and}\\
\beta_{p}&=\frac{3}{16 I_{0}(\kappa )}(\lambda^{3}-4\lambda )(I_{p+1}(\kappa )-I_{p-1}(\kappa ))-\frac{\lambda^{3}}{16 I_{0}(\kappa )}(I_{p+3}(\kappa )-I_{p-3}(\kappa )) .
%=\frac{I_{p}(\kappa )}{I_{0}(\kappa )}\left[ \left\{ -\frac{3p}{2\kappa^{2}}\frac{I_{p+1}(\kappa )}{I_{p}(\kappa )}+\frac{(p-2)(p-1)p}{2\kappa^{3}}\right\}\lambda^{3}+\frac{3p\lambda}{2\kappa}\right] .
\end{align*}
In addition, the $p$th cosine and sine moments for the ESS-vM distribution of order $2$ with $\mu =0$ are given by
\begin{align}
\alpha_{p}&=\frac{I_{p}(\kappa )}{I_{0}(\kappa )}\quad \textrm{and} \label{ESSvM2cos} \\
\beta_{p}&=\frac{15}{128}c_{2,1}(\lambda)\left( \frac{I_{p+1}(\kappa )-I_{p-1}(\kappa)}{I_{0}(\kappa)}\right) +\frac{5}{256}c_{2,2}(\lambda)\left( \frac{I_{p+3}(\kappa )-I_{p-3}(\kappa)}{I_{0}(\kappa)}\right) \notag \\
         &-\frac{3}{256}\lambda^{5}\left( \frac{I_{p+5}(\kappa )-I_{p-5}(\kappa)}{I_{0}(\kappa)}\right) ,\label{ESSvM2sin} 
\end{align}
where $c_{2,1}(\lambda )=-\lambda^{5}+4\lambda^{3}-8\lambda$ and $c_{2,2}(\lambda )=3\lambda^{5}-8\lambda^{3}$.
\end{example}

Subsequently,  we consider the ESS-WC distribution of order $m$ with density
\begin{equation}
f_{\textrm{ESSWC}}^{(m)}(\theta ;\bm{\eta})=\frac{1 - \rho^2}{\pi (1 + \rho^2 - 2 \rho \cos (\theta  - \mu ))}G_{m}(\lambda \sin (\theta -\mu)), \label{ESSWC1}
\end{equation}
where $\bm{\eta}=(\mu ,\rho ,\lambda )^T$, $-\pi\leq\mu <\pi$, $\rho > 0$, and $-1\leq \lambda \leq 1$.

\begin{example}\label{ex:moments2}
The ESS-WC distribution of order $1$ with $\mu =0$ has the cosine and sine moments as follows:
\begin{align*}
\alpha_{p}&=\rho^{|p|},\qquad \textrm{and}\\
\beta_{p}&=\frac{3}{16}(\lambda^{3}-4\lambda )(\rho^{|p+1|}-\rho^{|p-1|})-\frac{\lambda^{3}}{16}(\rho^{|p+3|}-\rho^{|p-3|}), \qquad (p\in\mathbb{Z}).
\end{align*}
In addition, the $p$th cosine and sine moments for the ESS-WC distribution of order $2$ with $\mu =0$ are given by
\begin{align}
\alpha_{p}&=\rho^{|p|},\qquad \textrm{and} \label{ESSWC2cosine}\\
\beta_{p}&=\frac{15}{128}c_{2,1}(\lambda)\left( \rho^{|p+1|}-\rho^{|p-1|}\right) +\frac{5}{256}c_{2,2}(\lambda)\left( \rho^{|p+3|}-\rho^{|p-3|}\right) \notag \\
         &-\frac{3}{256}\lambda^{5}\left(\rho^{|p+5|}-\rho^{|p-5|}\right) ,\label{ESSWC2sine}
\end{align}
where $c_{2,1}(\lambda )$ and $c_{2,2}(\lambda )$ are given in Example \ref{ex:moments1}.
\end{example}

\begin{figure}[htb]
\begin{minipage}{.49\linewidth}
\centering
\includegraphics[width=0.98\linewidth]{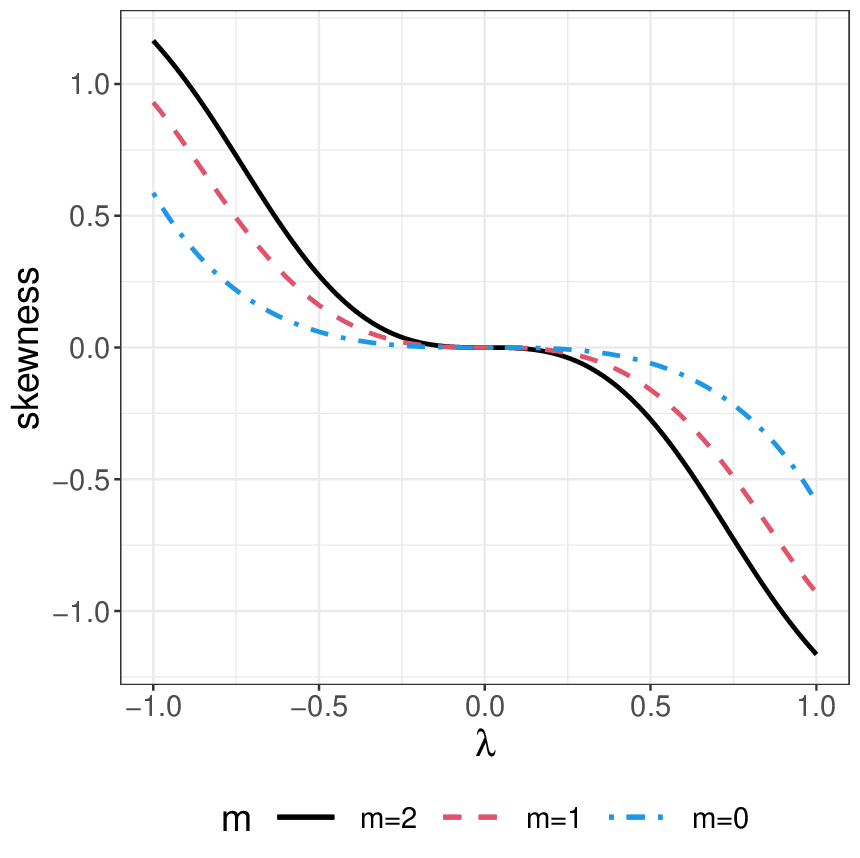}
\end{minipage}
\begin{minipage}{.49\linewidth}
\centering
\includegraphics[width=0.98\linewidth]{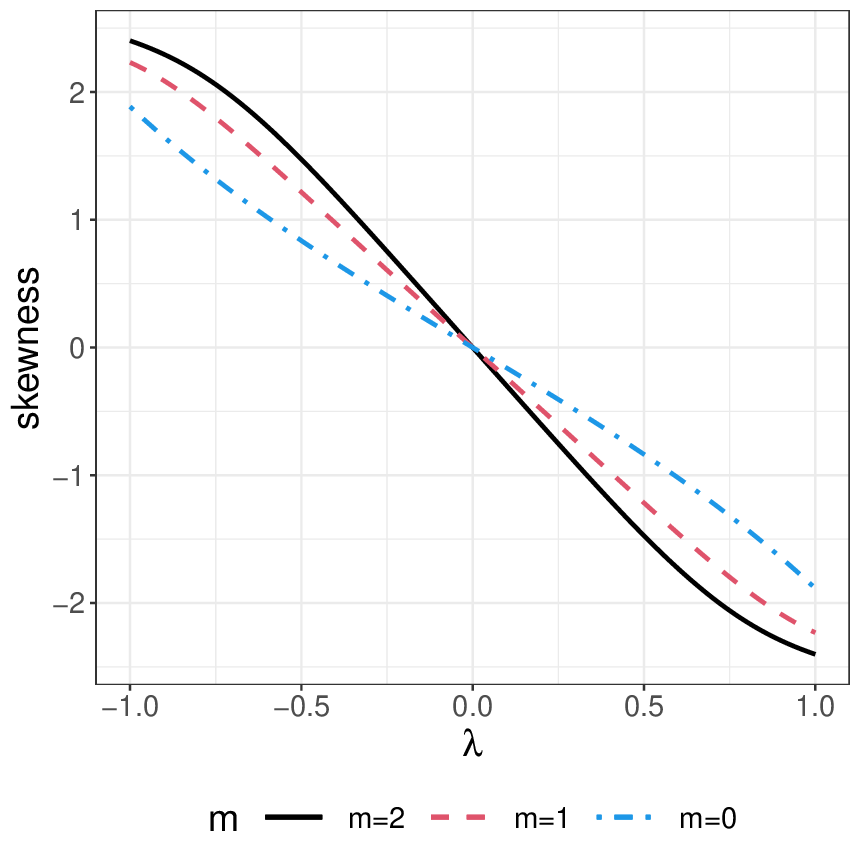}
\end{minipage}
\caption{Plots of the circular skewness measures for ESS-vM (left) and ESS-WC (right) distribution. Concentration parameters are chosen as $\kappa=2$ for ESS-vM distribution and $\rho=0.8$ for ESS-WC distribution. The location parameter is fixed at $\mu=0$ for both models.}
\label{skewness}
\end{figure}

For illustrative purposes, we show the circular skewness measure as a function of $\lambda$ for $m\in\{0,1,2\}$ with ESS-vM and ESS-WC distributions in Figure~\ref{skewness}. Here, concentration parameters are set at $\kappa=2$ and $\rho=0.8$ for ESS-vM and ESS-WC distributions, respectively. Recall that circular skewness (see, e.g. Chapter 2 of \citet{MJ09}) is defined as 
\begin{align*}
s=\frac{\bar{\beta_2}}{(1-\rho_{\textrm{MRL}}^{(1,m)})^{3/2}},
\end{align*}
where $\bar{\beta_2}:=E\left\{ \sin \left( 2(\Theta -\mu_{\textrm{MD}}^{(1,m)})\right)\right\}$ and $\mu_{\textrm{MD}}^{(1,m)}$ is the MD defined in Section \ref{Sec3.1}. 
%For details, see Chapter 2 of \citet{MJ09}. 
These figures show that as $m$ increases, the degree of the skewness also increases, which indicates that it can appropriately model the skewed structure of the data using an ESS distribution family.

Table \ref{table1} presents the range of skewness achievable by the ESS-vM distribution for $\mu=0$ and $m=0,1,2$. Notably, when $m=0$, the distribution represents the sine-skewed von Mises distribution. 

Comparison with Table 1 of \citet{BRAL22} reveals that each skewness range for the ESS-vM distributions with orders $1$ and $2$ is wider than the corresponding ranges for two generalized skew-symmetric von Mises distributions (GSSVM). This is because the largest interval for skewness in the two GSSVM distributions is $(-0.7878, 0.7878)$.
When the base density is the wrapped Cauchy distribution, the skewness that the ESS-WC distribution can attain goes to infinity as $\lambda=-1$ and $\rho \to 1$, and goes to minus infinity as $\lambda=1$ and $\rho \to 1$.

\begin{table}[htb]
\caption{Range of skewness of the ESS-vM distributions with $\mu=0$ and $m=0,1,2$}
\label{table1}
\begin{center}
\begin{tabular}{ll} 
ESSvM & Skewness \\\hline
$m=0$(sine-skewed) & $(-0.5875,0.5875)$ \\
$\kappa$ & 2.1342,2.1342 \\
$\lambda$ & 1, $-1$ \\
$m=1$ & $(-0.9301,0.9301)$ \\
$\kappa$ & 1.9661 \\
$\lambda$ & 1, $-1$ \\
$m=2$ & $(-1.1644,1.1644)$ \\
$\kappa$ & 2.0404,2.0404 \\
$\lambda$ & 1, $-1$ \\\hline
\end{tabular}
\end{center}
\end{table}

\subsection{Identifiability}
In this section, we show the identifiability for a family of the ESS-vM distributions and ESS-WC distributions. 
First, we consider the ESS-vM distribution of order $m$ with density
\begin{equation}
f_{\textrm{ESSvM}}^{(m)}(\theta ;\bm{\eta})=\frac{1}{\pi I_{0}(\kappa )}\exp\left\{ \kappa \cos (\theta -\mu )\right\}G_{m}(\lambda \sin (\theta -\mu)), \label{ESSWvM1}
\end{equation}
where $\bm{\eta}=(\mu ,\kappa ,\lambda )^T$. Let $\bm{H} =\{ \bm{\eta}|\mu \in [-\pi ,\pi ),\kappa >0, \lambda \in [-1,1]\}$ be a parameter space, and let $\mathcal{F}_{\textrm{vM}}^{(m)}$ be the family $\{ f_{\textrm{ESSvM}}^{(m)}(\theta ;\bm{\eta})|\bm{\eta}\in \bm{H} \}$.
\begin{proposition}\label{prop_identi1}
For any integer $m\geq 0$, the family $\mathcal{F}_{\textrm{vM}}^{(m)}$ is identifiable.
\end{proposition}
The proof is given in Appendix \ref{appendix}.

Next, we consider the ESS-WC distribution of order $m$ with density \eqref{ESSWC1}. Let $\bm{H} =\{ \bm{\eta}|\mu \in [-\pi ,\pi ),\rho \in (0,1), \lambda \in [-1,1]\}$ be a parameter space, and let $\mathcal{F}_{\textrm{WC}}^{(m)}$ be the family $\{ f_{\textrm{ESSWC}}^{(m)}(\theta ;\bm{\eta})|\bm{\eta}\in \bm{H} \}$. Then, the following result holds.

\begin{proposition}\label{prop_identi2}
For any integer $m\geq 0$, the family $\mathcal{F}_{\textrm{WC}}^{(m)}$ is identifiable. 
\end{proposition}
The proof is given in Appendix \ref{appendix}.
%We remark that it is complicated to provide identifiability conditions for von Mises base densities for general $m$.

\subsection{Random number generation}
In this section, we propose a random number generation method from the ESS circular distribution. 
To do that, we assume that the base density $f_0(\cdot ;\bm{\rho})$ satisfies Assumptions A1 and A2: 
\begin{itemize}
\setlength{\parskip}{0cm} 
\setlength{\itemsep}{0cm} 
\item[A1] The base density $f_0(\cdot ;\bm{\rho})$ is a continuous, unimodal, and reflective symmetric.
\item[A2] The first cosine moment of $f_0(\cdot ;\bm{\rho})$, $\alpha_{0,1}:=E_{0}\{ \cos (\Theta )\}$ takes a value in $(0,1)$. In addition, this is uniformly less than unity in $\bm{\rho}$, that is $\sup_{\rho}\alpha_{0,1}(\bm{\rho})<1$.
\end{itemize}
By the classical way to generate random numbers from Azzalini-type skew distributions, a random number is generated from the ESS circular density \eqref{ESScircular} with location $\mu =0$ by the following algorithm:
\begin{enumerate}
\setlength{\parskip}{0cm} 
\setlength{\itemsep}{0cm} 
\item Generate a random number $\Phi$ from the base density $f_{0}(\theta ;\bm{\rho})$.
\item Generate a random number $U$, which is independent of $\Phi$, from the
uniform distribution on the support $[0,1]$.
\item Set
\begin{align}
\Theta =\begin{cases}
\Phi \quad \text{if } U < G_m(\lambda \sin \Phi),
\\
-\Phi \quad \text{if } U \geq G_m(\lambda \sin \Phi).
\end{cases}
\label{rn}
\end{align}
\end{enumerate}
For the case with $\mu \ne 0$, setting $\tilde{\Theta}=(\Theta +\mu +\pi ) (\textrm{mod}2\pi) -\pi$ for $\Theta$ generated by Equation \eqref{rn}, the random variable $\tilde{\Theta}$ has the ESS circular density \eqref{ESScircular} with $\mu\ne 0$.

\begin{figure}[htb]
\begin{minipage}{.49\linewidth}
\centering
\includegraphics[width=0.98\linewidth]{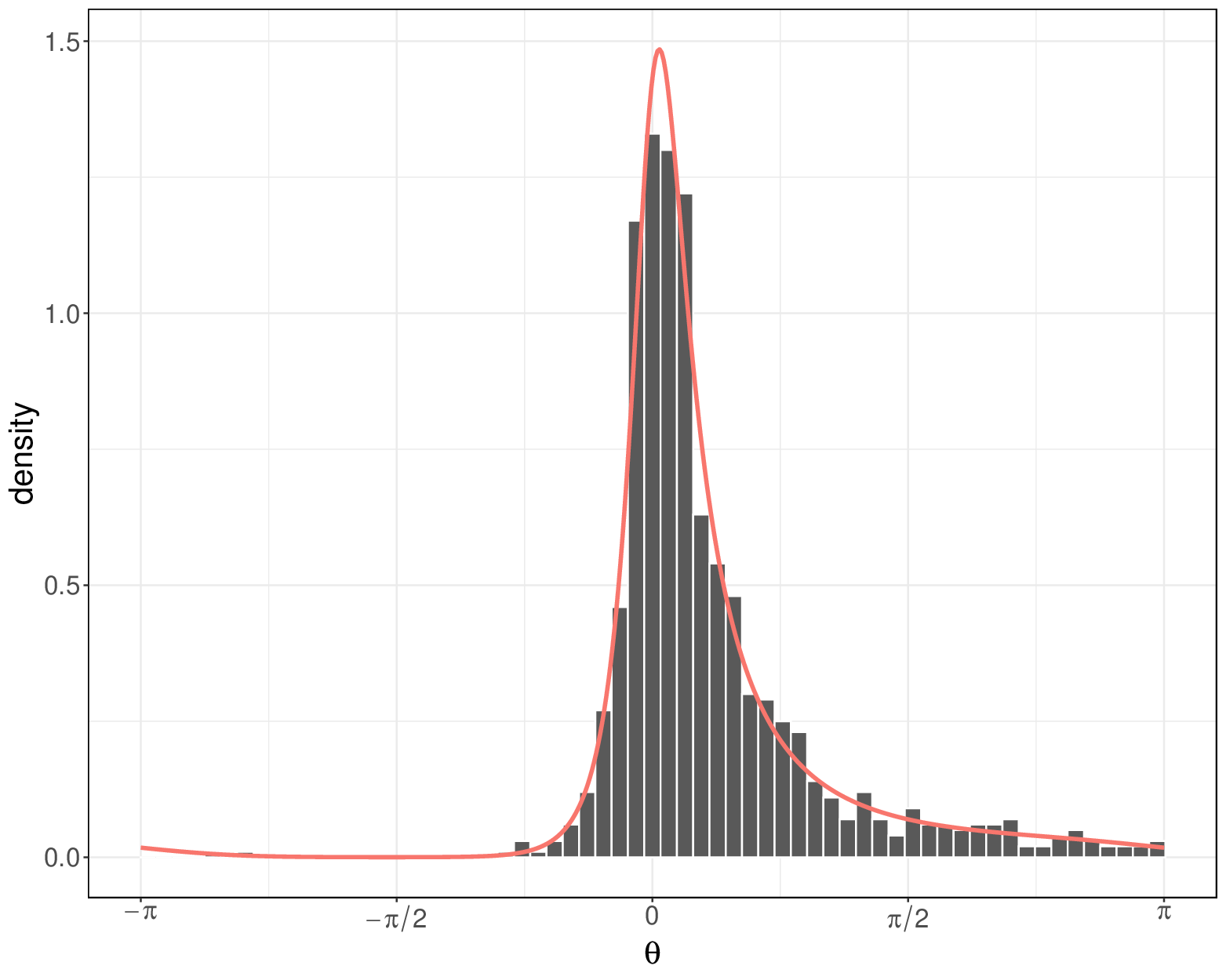}
\end{minipage}
\begin{minipage}{.49\linewidth}
\centering
\includegraphics[width=0.98\linewidth]{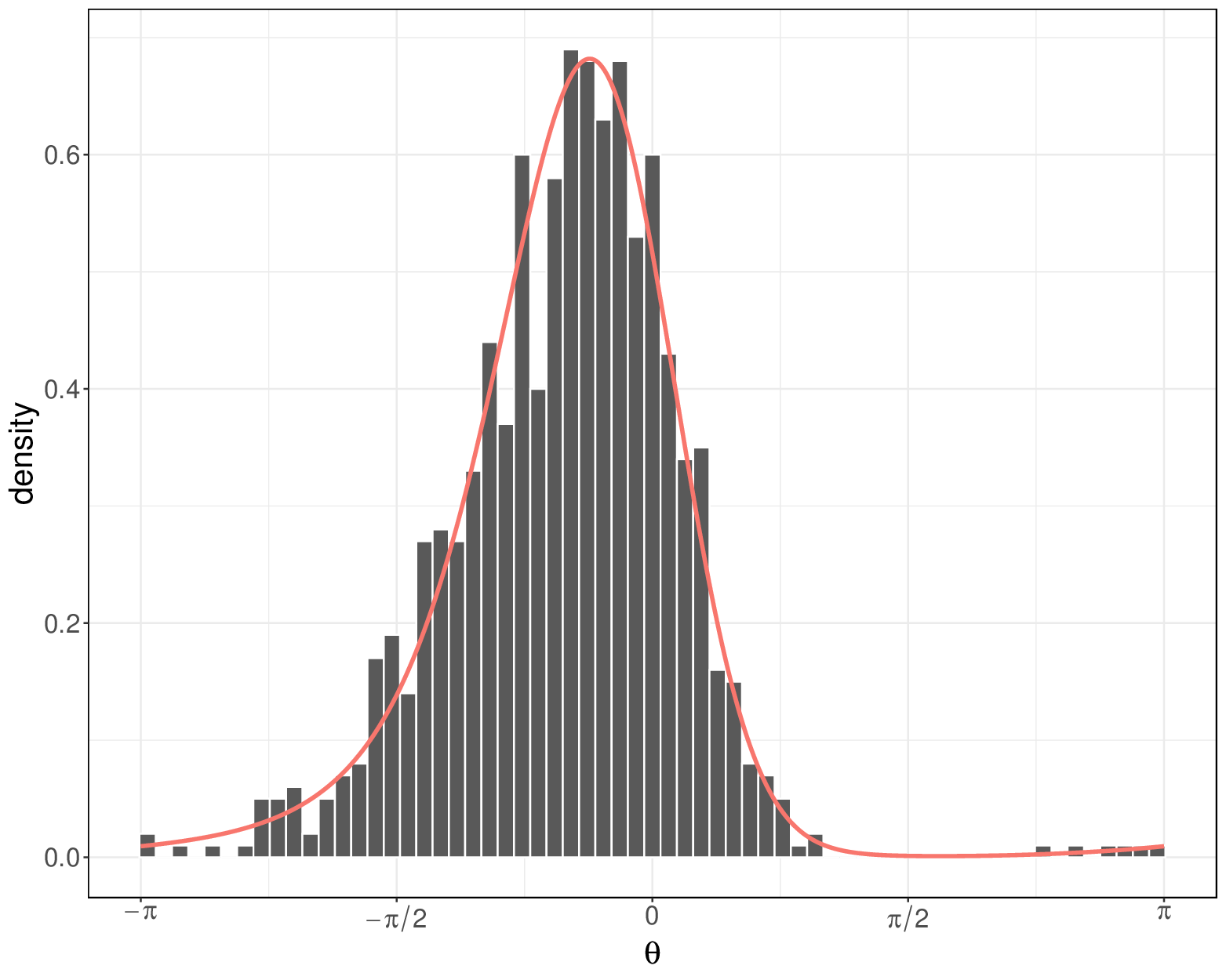}
\end{minipage}
\caption{Histograms of $n=1000$ random samples from ESS-WC distribution with parameters $\mu=0$, $\rho=0.8$, $\lambda=0.8$, and $m=3$ (left), and from ESS-vM distribution with parameters $\mu=0$, $\kappa=2$, $\lambda=-0.8$, and $m=2$ (right).}
\label{fig5}
\end{figure}

Figure~\ref{fig5} shows the histograms of the generated random numbers with ESS-WC and ESS-vM distributions together with corresponding density functions. The parameters for ESS-WC distributions are chosen as $\mu=0$, $\rho=0.8$, $\lambda=0.8$, and for ESS-vM distributions as $m=3$ and  $\mu=0$, $\kappa=2$, $\lambda=-0.8$, and $m=2$.

\section{Parameter estimation}\label{Sec4}
We now use the maximum likelihood method to estimate the unknown model parameters. 
Let $\bm{\theta}_n=(\theta_1, \ldots, \theta_n)$ be a random sample of size $n$ from the ESS circular density $f_{\textrm{ESS}}^{(m)}(\theta ;\mu ,\bm{\rho},\lambda )$ of order $m$, where $m$ is a prespecified constant.
Denote a parameter vector by $\bm{\eta}=(\mu, \bm{\rho}, \lambda )^T$. 
Suppose that $M\subseteq [-\pi ,\pi )$, $\bm{R}$, and $\Lambda :=[-1+\delta_{\lambda},1-\delta_{\lambda}]$ are parameter spaces of $\mu$, $\bm{\rho}$, and $\lambda$, respectively, where $\delta_{\lambda}>0$ is a small positive constant.
Then, we define a parameter space of $\bm{\eta}=(\mu ,\bm{\rho},\lambda )^{T}$ as $\bm{H}=M\times \bm{R}\times \Lambda$. 
The log-likelihood function of $\bm{\eta}$ based on the observed sample is written as 
\begin{align}
\ell_m(\bm{\eta}) = \frac{1}{n}\sum_{i=1}^n \left[ 
\log(f_0(\theta_i-\mu ;\bm{\rho}))+\log\{ 2G_m(\lambda \sin(\theta_i-\mu))\}
\right] .\label{log-likeli}
\end{align}
Then, the maximum likelihood estimator (MLE) $\hat{\bm{\eta}}_n$ satisfies that $\ell_m(\hat{\bm{\eta}}_n)\geq \ell_m(\bm{\eta})$ for any $\bm{\eta}\in \bm{H}$. 

In addition to conditions A1 and A2, we assume the following conditions:
\begin{itemize}
\setlength{\parskip}{0cm} 
\setlength{\itemsep}{0cm} 
\item[A3] 
\begin{itemize}
\item[(i)] $\bm{H}$ is compact, and the true parameter vector $\bm{\eta}_0$ belongs to $\bm{H}$.
\item[(ii)] $\bm{H}$ is convex and compact, and the true parameter vector $\bm{\eta}_0$ belongs to the interior of $\bm{H}$. 
\end{itemize}
\item[A4] 
There exist constants $Q>0$ and $q>0$ such that for any $\theta\in[-\pi,\pi)$, 
\begin{align}
\sup_{\bm{\rho}\in \bm{R}}f_{0}(\theta ; \bm{\rho}) <Q ~~\text{and}~~
\inf_{\bm{\rho}\in \bm{R}}f_{0}(\theta ; \bm{\rho}) >q. \label{condA4}
\end{align}
\item[A5] The first three derivatives of base density $f_0(\theta-\mu;\bm{\rho})$ with respect to parameters $\mu$ and $\bm{\rho}$ are continuous and bounded for all $\theta$ and $\bm{\eta} \in \bm{H}$.
\item[A6] The Fisher information matrix $\bm{I}(\bm{\eta}):=-E_{0}\left\{ (\partial^{2}/\partial \bm{\eta}\partial \bm{\eta}^{T})\log f_{\textrm{ESS}}^{(m)}(\Theta ;\bm{\eta})\right\}$ is non-singular.
\item[A7] The family $\{ f_{\textrm{ESS}}^{(m)}(\theta ;\bm{\eta})|\bm{\eta}\in\bm{H}\}$ is identifiable, that is, $\bm{\eta}_1 \ne \bm{\eta}_2$ implies $f_{\textrm{ESS}}^{(m)}(\theta ;\bm{\eta}_1)\ne f_{\textrm{ESS}}^{(m)}(\theta ;\bm{\eta}_2)$.
\end{itemize}
To ensure the compactness of $\bm{H}$ in condition A3, we need to restrict the parameter space of $\mu$ to $[-\pi,\pi -\delta_{\mu}]$ with a small constant $\delta_{\mu}>0$. However, this can be relaxed to the condition $[-\pi,\pi )$ by the approach of \citet{AISM21}. 
The first inequality in the condition \eqref{condA4} ensures the boundedness condition, that is $E_{\bm{\eta}_{0}}\left\{ \sup_{\bm{\eta}\in U}\log f_{\textrm{ESS}}^{(m)}(\Theta ;\bm{\eta}) \right\} < \infty$ for every neighbourhood $U$.Whereas this inequality holds for the von Mises distribution and wrapped Cauchy distribution, it does not hold for the cardioid distribution.
Hence, condition A4 is slightly stronger than the foregoing boundedness condition.
The second inequality in expression \eqref{condA4} and the parameter space $\Lambda :=[-1+\delta_{\lambda},1-\delta_{\lambda}]$ guarantee the integrability condition $E_{\bm{\eta}_{0}}\left\{ \log f_{\textrm{ESS}}^{(m)}(\Theta ;\bm{\eta}) \right\} >-\infty$ for any $\bm{\eta}$ because it holds that $E_{\bm{\eta}_{0}}\left\{ \log G_{m}(\lambda \sin (\Theta -\mu ))\right\} >-\infty$ for any $\lambda\in\Lambda$ and $\mu \in M$. 
If we only show the consistency of the maximum likelihood estimators, the parameter space of $\lambda$ might be extended to $[-1,1]$. We omit the details here, but the interested reader can refer to Remark 1 and Proposition 1 of \citet{MSA20}. 
To the best of the authors' knowledge, no SRS circular model including the proposed model satisfies condition A6 theoretically. 
However, for the ESS-WC distribution, the Hessian matrix of minus the log-likelihood function, which is an estimate of the Fisher information matrix, is numerically non-degenerate. In contrast, the Fisher information matrix in the ESS-vM distribution becomes singular at the point $\lambda =0$. This is pointed out in \citet[p.71]{LV17a}. 
Propositions \ref{prop_identi1} and \ref{prop_identi2} show that Condition A7 holds for the ESS-WC and ESS-vM distributions of general order $m$.

We obtain the following results for consistency and asymptotic normality of the MLE.

\begin{theorem}\label{consistency_AN}
Suppose that conditions A1, A2, A3(i), A4, and A7 hold. If the base density $f_{0}(\theta -\mu ; \bm{\rho})$ is continuous in $\mu$ and $\bm{\rho}$, it follows that
\begin{align*}
\hat{\bm{\eta}}_n \mathop{\longrightarrow}_{p} \bm{\eta}_0 ~~~as ~~~n\to \infty .
\end{align*}

Suppose that conditions A1, A2, A3(ii), A4--A7 hold. Then, it follows that
\begin{align}
\sqrt{n}( \hat{\bm{\eta}}_n - \bm{\eta}_0 ) \mathop{\longrightarrow}_{d} N (\bm{0}, \bm{I}(\bm{\eta}_0)^{-1})~~~as ~~~n\to \infty, \label{AN}
\end{align}
where $\mathop{\to}_{p}$ and $\mathop{\to}_{d}$ represent the convergence in probability and the convergence in distribution, respectively.
\end{theorem}

\section{Simulations}\label{Sims}

We now investigate the performances of the finite sample estimations via Monte Carlo simulations. We choose two models from the ESS family: ESS-vM and ESS-WC distributions. For the parameters of the ESS-vM distribution, we set $\mu=0$, $\kappa=2$, $m=2$, and the skewness parameter $\lambda$ has three cases with $\lambda \in \{-0.2, -0.5, -0.8\}$. Similarly, the parameters of the ESS-WC distribution are set at $\mu=0$, $\rho=0.8$, $m=3$, and $\lambda\in \{0.2, 0.5, 0.8\}$. For each case, sample sizes are set at $n=100, 200$, and $n=500$ to verify the consistency and unbiasedness of the estimators. The number of replicates for the simulations is 1000, and we obtain average estimates of the MLEs together with their root mean squared errors (RMSE). 

Table \ref{tab:sim1} summarizes the simulation results of the MLEs. As the sample size increases, the RMSE decreases for all cases, indicating the consistency of the estimators. As for the bias of the estimates, the skewness parameter $\lambda$ tends to underestimate and overestimate compared with its negative and positive true values, respectively, whereas these biases decrease as the sample size increases. 

% Sun Mar 20 11:40:48 2022
\begin{landscape}
\begin{table}[ht]
\centering
\caption{Averages and RMSEs for the MLEs of simulated samples.}
\label{tab:sim1}
\begin{tabular}{lrrrrrrrrr}
  \hline\hline
   & \multicolumn{3}{l}{$n=100$} & \multicolumn{3}{l}{$n=200$}  &
   \multicolumn{3}{l}{$n=500$} \\ \hline
  &  \multicolumn{9}{l}{ESS-vM distribution with $m=2$} \\
 & $\mu(=0)$ & $\kappa(=2)$ & $\lambda$ & $\mu(=0)$ & $\kappa(=2)$ & $\lambda$ & $\mu(=0)$ & $\kappa(=2)$ & $\lambda$ \\  
  \hline
Mean:$(\lambda=-0.2)$ & 0.0003 & 1.9583 & $-$0.2062 & 0.0003 & 1.9535 & $-$0.2008 & $-$0.0058 & 1.9719 & $-$0.1880\\ 
RMSE: & 0.1182 & 0.3413 & 0.2479 & 0.1025 & 0.2526 & 0.2143 & 0.0800 & 0.1692 & 0.1685 \\ 
Mean:$(\lambda=-0.5)$ & $-$0.0437 & 2.1271 & $-$0.4033 & $-$0.0305 & 2.0776 & $-$0.4289 & $-$0.0134 & 2.0340 & $-$0.4680 \\ 
RMSE:& 0.1258 & 0.4528 & 0.3071 & 0.0978 & 0.3349 & 0.2382 & 0.0669 & 0.2206 & 0.1598 \\ 
Mean:$(\lambda=-0.8)$ & $-$0.0264 & 2.1531 & $-$0.7493 & $-$0.0078 & 2.0483 & $-$0.7924 &  0.0001 & 2.0094 & $-$0.8094  \\ 
RMSE: & 0.0948 & 0.4925 & 0.2828 & 0.0580 & 0.3120 & 0.1723 & 0.0260 & 0.1735 & 0.0844 \\
\hline
 &  \multicolumn{9}{l}{ESS-WC distribution with $m=3$} \\
 & $\mu(=0)$ & $\rho(=0.8)$ & $\lambda$ & $\mu(=0)$ & $\rho(=0.8)$ & $\lambda$ 
 & $\mu(=0)$ & $\rho(=0.8)$ & $\lambda$ \\ 
Mean:$(\lambda=0.2)$  & 0.0000 & 0.7996 & 0.2078 & $-$0.0007 & 0.7997 & 0.2089 & $-$0.0001 & 0.7997 & 0.2017 \\
RMSE: & 0.0208 & 0.0267 & 0.1592 & 0.0142 & 0.0177 & 0.1032 & 0.0087 & 0.0116 & 0.0609 \\ 
Mean:$(\lambda=0.5)$ & $-$0.0036 & 0.8002 & 0.5424 & $-$0.0010 & 0.7993 & 0.5188 & $-$0.0008 & 0.8000 & 0.5085 \\ 
RMSE:&  0.0222 & 0.0286 & 0.1908 & 0.0149 & 0.0188 & 0.1231 & 0.0086 & 0.0120 & 0.0686 \\ 
Mean:$(\lambda=0.8)$ & $-$0.0017 & 0.7980 & 0.8334 & $-$0.0020 & 0.7987 & 0.8320 & $-$0.0016 & 0.7990 & 0.8239 \\ 
RMSE: &   0.0192 & 0.0280 & 0.1612 & 0.0138 & 0.0196 & 0.1326 & 0.0092 & 0.0134 & 0.0957 \\ 

   \hline
\end{tabular}
\end{table}
\end{landscape}

In the previous simulations, we assume that the underlying base density $f_0$ and the order of the model are known in advance. We also investigate for estimating order parameter $m$ for the possible candidate sets of $m \in\{0,1,2,3,4\}$ using the AIC and Takeuchi information criterion (TIC). Denote $F$ and $F'$ be the distribution for fitting with density $f_{\textrm{ESS}}^{(m)}(\theta|\bm{\eta})$ and true distribution, respectively. The definitions for the AIC and TIC are as follows:
\begin{align*}
\textrm{AIC} = -2 \ell_m(\hat{\bm{\eta}}_n)+ 2p
\end{align*}
and
\begin{align*}
\textrm{TIC} = -2 \ell_m(\hat{\bm{\eta}}_n)+ 2 %\textrm{bias} ~~
%\textrm{bias} =
 \textrm{tr} \{ \hat{\bm{J}}^{-1}(F') \hat{\bm{I}}(F')\},
\end{align*}
respectively. Here $\ell_m$ is the log-likelihood defined in Equation \eqref{log-likeli}, and $\hat{\bm{J}}(F')$ and $\hat{\bm{I}}(F')$ are the consistent estimators of 
\begin{align*}
\bm{J}(F') = -E_{F'}\left[
\frac{\partial^2 \log f_{\textrm{ESS}}^{(m)}(\Theta|\bm{\eta})}{\partial \bm{\eta}\partial \bm{\eta}^T}
\right]~~\textrm{and}~~
\bm{I}(F') = E_{F'}\left[
\frac{\partial \log f_{\textrm{ESS}}^{(m)}(\Theta|\bm{\eta})}{\partial \bm{\eta}}
\frac{\partial \log f_{\textrm{ESS}}^{(m)}(\Theta|\bm{\eta})}{\partial \bm{\eta}^T}
\right],
\end{align*}
where expectations are evaluated under the true distribution $F'$. We call the term $\textrm{tr} \{ \hat{\bm{J}}^{-1}(F') \hat{\bm{I}}(F')\}$ the ``bias" or ``penalty," and notice that $\bm{I}(F)$ corresponds to the Fisher information matrix. Recall that the number of the parameters among the models are the same, so $p=3$. The evaluating expectations under the density $F'=F$ that indicates the estimating model $F'$ are included in the true model $F$ yielding $\bm{J}(F)=\bm{I}(F)$, which reduces bias terms being 3. See, \citet[Chapter~3]{konishi2008information} for details. 

Tables~\ref{tabIC1} and \ref{tabIC2} summarize the results of the selected order $\hat{m}^{\textrm{(MLL)}}:=\textrm{argmax}_{m\in\{0,1,2,3,4\}} \ell_m(\hat{\bm{\eta}}_n)$ and those obtained by minimizing TIC for the ESS-vM and ESS-WC distributions, respectively. Recall that the number of the parameters is the same among models; the estimated order $\hat{m}^{\textrm{(MLL)}}$ is equal to that minimizing AIC. Hence, we compare the order selection between the AIC and TIC. The numbers of the true model order $m$ and their corresponding frequencies of the selected models are presented in boldface. For both cases, the AIC tends to select larger order $\hat{m}$ than that by using TIC. The performances between the AIC and TIC of the rates of the true order selection are almost identical for the ESS-vM distribution, whereas the TIC outperforms AIC for the ESS-vM distribution with $n=100$ and $200$. This is due to the differences in the concentration of the models; simulated ESS-vM distribution is much more concentrated than the ESS-vM distribution, in which we find, we found that the TIC works better than AIC for small samples.

 \begin{table}[ht]
\centering
\caption{Results for the model selection with AIC and TIC for the ESS-vM distribution with $m=2$.}
\label{tabIC1}
\begin{tabular}{lrrrrrrrrrr}
  \hline \hline
      &\multicolumn{5}{c}{$\hat{m}^{\textrm{(MLL)}}$} & \multicolumn{5}{c}{TIC} \\ 
         $m$& 0 & 1 & $\bm{2}$ & 3 & 4 & 0 & 1 & $\bm{2}$ & 3 & 4 \\ 
         \hline
      \multicolumn{11}{l}{$\lambda=-0.2$} \\  
$n=100$  & 257 & 117 &  $\bm{76}$ & 102 & 448 & 426 & 107 & $\bm{107}$ & 95 & 265 \\ 
$n=200$  & 250 & 95 &  $\bm{83}$ & 85 & 487 & 440 & 98 &  $\bm{101}$ & 79 & 282 \\ 
$n=500$  & 302 & 76 &  $\bm{85}$ & 73 & 464 & 444 & 105 &  $\bm{77}$ & 88 & 286 \\ 
      \multicolumn{11}{l}{$\lambda=-0.5$} \\  
$n=100$  & 270 & 125 &  $\bm{50}$ & 54 & 501 & 486 & 101 &  $\bm{62}$ & 55 & 296 \\ 
$n=200$  & 283 & 91 &  $\bm{61}$ & 39 & 526 & 421 & 75 &  $\bm{47}$ & 50 & 407 \\ 
$n=500$  & 306 & 104 &  $\bm{39}$ & 38 & 513 & 343 & 101 &  $\bm{52}$ & 51 & 453 \\ 
     \multicolumn{11}{l}{$\lambda=-0.8$} \\  
$n=100$  & 178 & 240 &  $\bm{137}$ & 85 & 360 & 339 & 259 & $\bm{133}$ & 67 & 202 \\  
$n=200$  & 155 & 297 &  $\bm{130}$ & 55 & 363 & 310 & 304 &  $\bm{96}$ & 28 & 262 \\ 
$n=500$  & 105 & 332 &  $\bm{162}$ & 65 & 336 & 202 & 295 &  $\bm{92}$ & 58 & 353 \\ 
   \hline
 \end{tabular}
\end{table}
   
    \begin{table}[htb]
\centering
\caption{Results for the model selection with AIC and TIC for the ESS-WC distribution with $m=3$.}
\label{tabIC2}
\begin{tabular}{lrrrrrrrrrr}
  \hline \hline
      &\multicolumn{5}{c}{$\hat{m}^{\textrm{(MLL)}}$} & \multicolumn{5}{c}{TIC} \\ 
      \hline 
    $m$& 0 & 1 & 2 & $\bm{3}$ & 4 & 0 & 1 & 2 & $\bm{3}$ & 4 \\ 
  \hline
    \multicolumn{11}{l}{$\lambda=0.2$} \\  
$n=100$  & 522 & 6 & 3 &  $\bm{2}$  & 467 & 442 & 86 & 82 &  $\bm{142}$  & 248 \\ 
$n=200$  & 528 & 13 & 2 &  $\bm{1}$  & 456 & 407 & 86 & 99 &  $\bm{139}$  & 269 \\ 
$n=500$   & 492 & 9 & 1 &  $\bm{1}$  & 497 & 404 & 85 & 113 &  $\bm{144}$  & 254 \\ 
      \multicolumn{11}{l}{$\lambda=0.5$} \\  
$n=100$  & 398 & 125 & 44 &  $\bm{20}$  & 413 & 503 & 127 & 59 &  $\bm{56}$  & 255 \\ 
$n=200$  & 419 & 108 & 25 &  $\bm{10}$  & 438 & 500 & 57 & 22 &  $\bm{40}$  & 381 \\ 
$n=500$ & 383 & 111 & 40 &  $\bm{20}$  & 446 & 380 & 93 & 19 &  $\bm{39}$  & 469 \\ 
      \multicolumn{11}{l}{$\lambda=0.8$} \\  
$n=100$ & 112 & 237 & 140 &  $\bm{89}$  & 422 & 272 & 244 & 129 &  $\bm{94}$  & 261 \\ 
$n=200$ & 69 & 246 & 212 &  $\bm{87}$  & 386 & 213 & 332 & 184 &  $\bm{84}$  & 187 \\ 
$n=500$ & 8 & 264 & 274 &  $\bm{97}$  & 357 & 39 & 536 & 147 &  $\bm{68}$  & 210 \\  
   \hline
\end{tabular}
\end{table}

\begin{figure}[htb]
\begin{minipage}{.49\linewidth}
\centering
\includegraphics[width=0.98\linewidth]{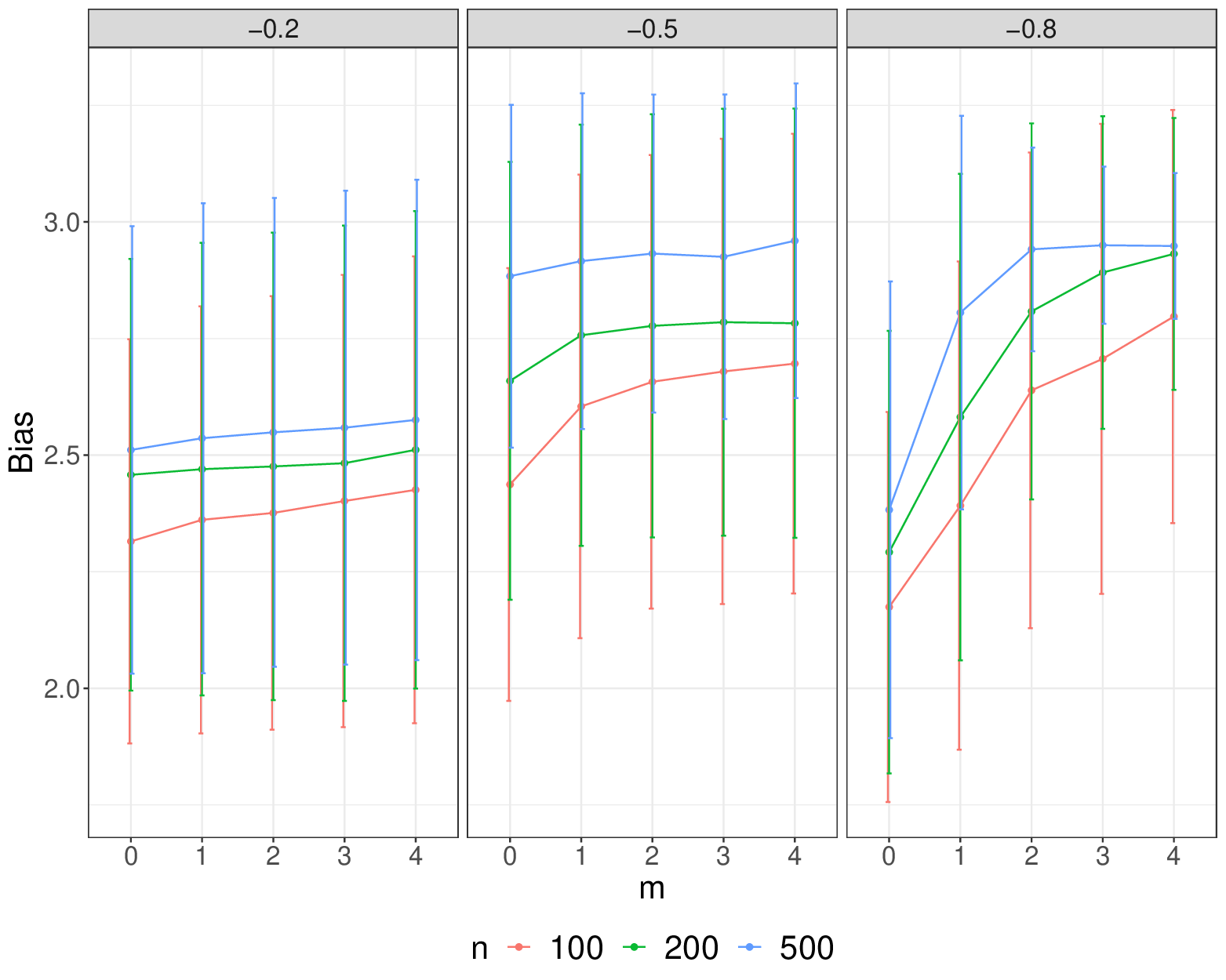}
\end{minipage}
\begin{minipage}{.49\linewidth}
\centering
\includegraphics[width=0.98\linewidth]{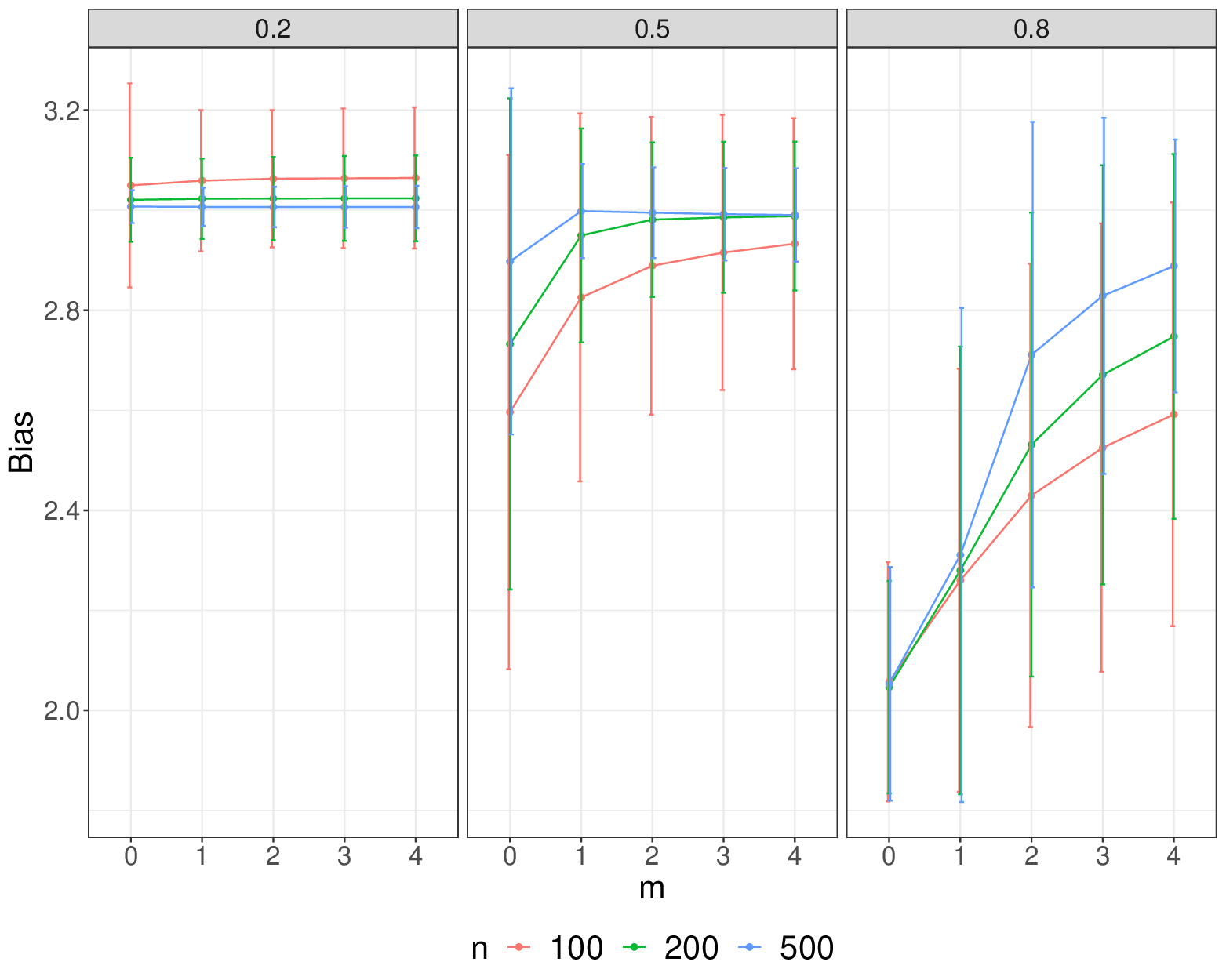}
\end{minipage}
\caption{The penalty term of the TIC as a function of $m$ for ESS-vM distribution with true $m=2$ (left) and for ESS-WC distribution with true $m=3$ (right).}
\label{fig:aic}
\end{figure}

Figure \ref{fig:aic} plots the average bias term of the TIC 
for ESS-vM and ESS-WC distributions, respectively. Notice that the bias term of the AIC is 3. The vertical lines in the figure indicate the standard deviations of the bias terms. As $m$ increases, the bias terms also increase, indicating that the TIC tends to select smaller order $m$ compared to the AIC. As $n$ increases, the bias term tends to converge to 3, which is the bias of the AIC. Hence, we conclude that estimating unknown order parameters by comparing both AIC and TIC is rather difficult, whereas small values of model order $m$ are achieved via TIC.

Table~\ref{tabBoundaryVM} summarizes the rate of the estimated skewness parameter $\lambda$ on the boundary of its parameter space, which is the fraction $1/1000 \sum_{i=1}^n I_{\{ |\hat{\lambda}_n|> 0.99\}}$. Here, $I_{\{\cdot \}}$ denotes the indicator function. This table shows that the absolute value of the true skewness parameter $\lambda$ approaches 1, and the rate of the boundary solutions increases. For the ESS models, as $m$ and $n$ increase, the rate of the boundary solutions decreases. The ESS-WC distributions have more boundary solutions than ESS-vM distributions, due to the larger concentration parameter chosen for the former. These findings confirm that the proposed ESS distributions have a high capability to treat the skewness of the observed data structures.

\begin{table}[htb]
\centering
\caption{The rate of the estimated skewness parameters laid around the boundary of the parameter space.}
\label{tabBoundaryVM}
\begin{tabular}{lrrrrrrrrrr}
  \hline \hline
  & \multicolumn{5}{c}{ESS-vM} & \multicolumn{5}{c}{ESS-WC} \\
 & 0 & 1 & 2 & 3 & 4 & 0 & 1 & 2 & 3 & 4 \\ 
  \hline
 &\multicolumn{5}{l}{$\lambda=-0.2$} &\multicolumn{5}{l}{$\lambda=0.2$} \\ 
 $n=100$ & 0.04 & 0.01 & 0.00 & 0.00 & 0.00 & 0.03 & 0.00 & 0.00 & 0.00 & 0.00 \\ 
 $n=200$ & 0.01 & 0.00 & 0.00 & 0.00 & 0.00 & 0.00 & 0.00 & 0.00 & 0.00 & 0.00 \\ 
 $n=500$ & 0.00 & 0.00 & 0.00 & 0.00 & 0.00 & 0.00 & 0.00 & 0.00 & 0.00 & 0.00 \\ 
 &\multicolumn{5}{l}{$\lambda=-0.5$}  &\multicolumn{5}{l}{$\lambda=0.5$}\\ 
 $n=100$ & 0.19 & 0.04 & 0.01 & 0.00 & 0.00 & 0.41 & 0.12 & 0.06 & 0.03 & 0.01\\ 
 $n=200$ & 0.09 & 0.00 & 0.00 & 0.00 & 0.00 & 0.29 & 0.03 & 0.00 & 0.00 & 0.00  \\ 
 $n=500$ & 0.01 & 0.00 & 0.00 & 0.00 & 0.00 & 0.12 & 0.00 & 0.00 & 0.00 & 0.00 \\
 &\multicolumn{5}{l}{$\lambda=-0.8$}  &\multicolumn{5}{l}{$\lambda=0.8$}\\  
 $n=100$ & 0.71 & 0.44 & 0.21 & 0.12 & 0.06 & 0.95 & 0.69 & 0.46 & 0.33 & 0.24 \\ 
 $n=200$ & 0.68 & 0.35 & 0.10 & 0.03 & 0.01 & 0.96 & 0.69 & 0.39 & 0.21 & 0.12\\ 
 $n=500$ & 0.68 & 0.19 & 0.01 & 0.00 & 0.00 & 0.96 & 0.71 & 0.24 & 0.09 & 0.02 \\ 
   \hline
\end{tabular}
\end{table}

\section{Real data examples}\label{Sec6}

To verify the performance of the proposed model, we use the following well-known datasets in circular statistics:
\begin{itemize}
\setlength{\parskip}{0cm} 
\setlength{\itemsep}{0cm} 
\item \textbf{Dataset 1}: Compass angle of the orientations of termite mounds of Amitermes Laurens (\#8 of \texttt{fisherB13} datasets; sample size: $n=48$)
\item \textbf{Dataset 2}: Compass angle of the orientations of termite mounds of Amitermes Laurens (\#1 of \texttt{fisherB13} datasets; sample size: $n=100$)
%\item \textbf{Dataset 3}: Compass angle of the orientations of cross-bedding structures and of the long axes of particles in undeformed sediments. The sample size is $n=580$.
\end{itemize}
This dataset is measured at 14 sites in Cape York Peninsula, North Queensland, and is available from \citet{Fi93} and the circular library in the \texttt{R} software. The angle measured in degrees is converted into radian, and $\pi$ is subtracted from the original data to make it easier to see the fitted probability density function here. 

The sample mean direction $\bar{\mu}$, the sample mean resultant length $\bar{R}$, and sample skewness $\hat{s}$ for these three datasets are, respectively as follows:
\begin{align*}
&\textrm{Dataset~1:}~~\bar{\mu}=-0.0989, ~~~\bar{R}=0.9427, ~~~\hat{s}=-0.0989,\\
&\textrm{Dataset~2:}~~\bar{\mu}=0.0489,~~~ \bar{R}=0.8826, ~~~\hat{s}=-1.3500.
%&\textrm{Dataset~3:}~~\bar{\mu}=-0.2977, ~~~\bar{R}=0.6607, ~~~\hat{s}=-2.9304.
\end{align*} 
The degree of skewness is more apparent for Dataset~2, while it is low for Dataset~1. We test the null hypothesis of reflective symmetry against the alternative of a skewed distribution \citep{pewsey2002testing}. The $p$-values of the test statistics for circular reflective symmetry for Datasets 1 and 2 are 0.9355 and 0.0889, respectively. Hence we observe that Dataset 2 has a negatively skewed distribution at the 10\%  significant levels, whereas we cannot find evidence of the asymmetry for Dataset 1.

\begin{figure}[htb]
		\centering\includegraphics[height=8cm,width=.9\textwidth]{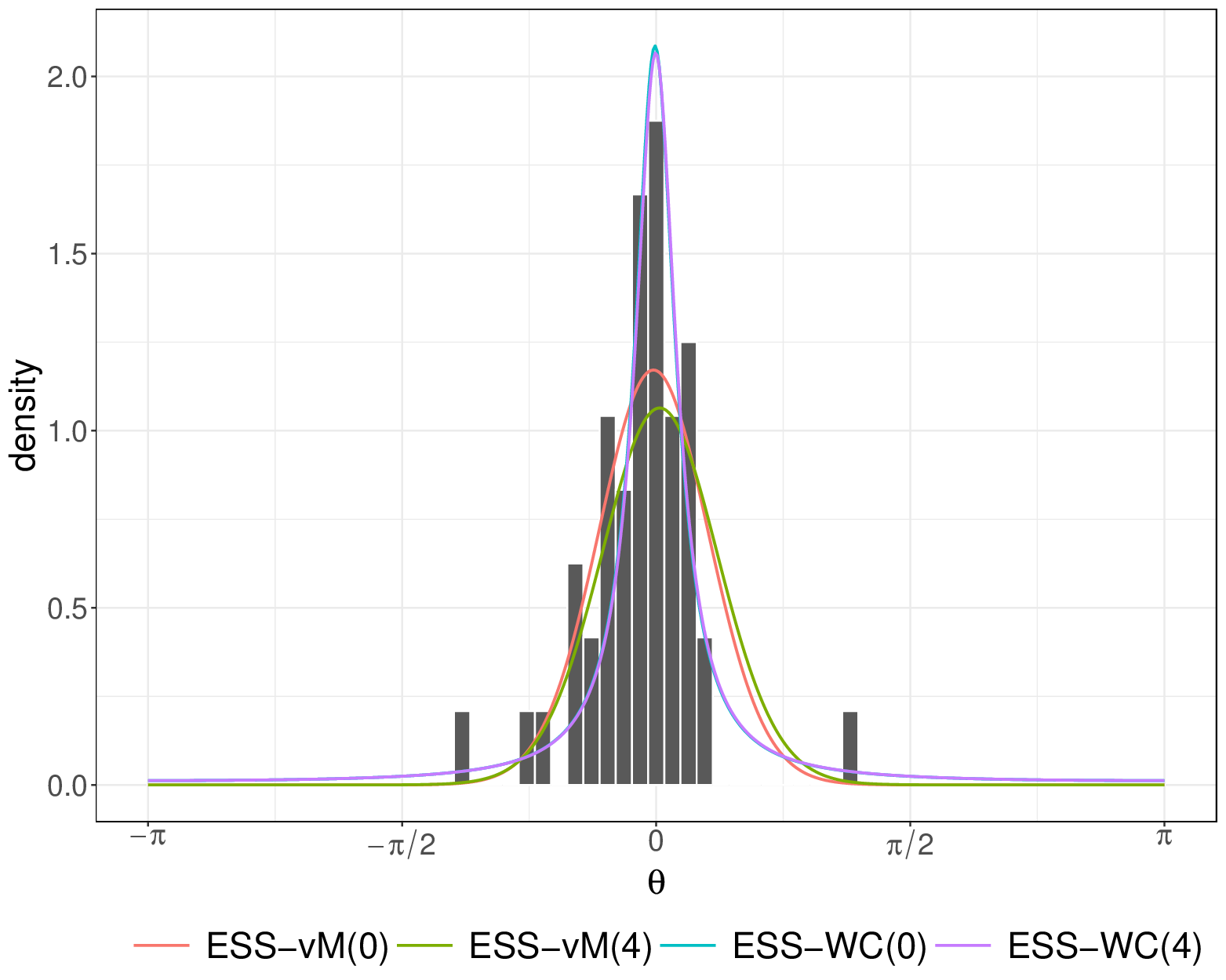}
		\caption{Histogram of dataset 1 with estimated density functions }
\label{hist_fig1}
\end{figure}

\begin{figure}[htb]
		\centering\includegraphics[height=8cm,width=.9\textwidth]{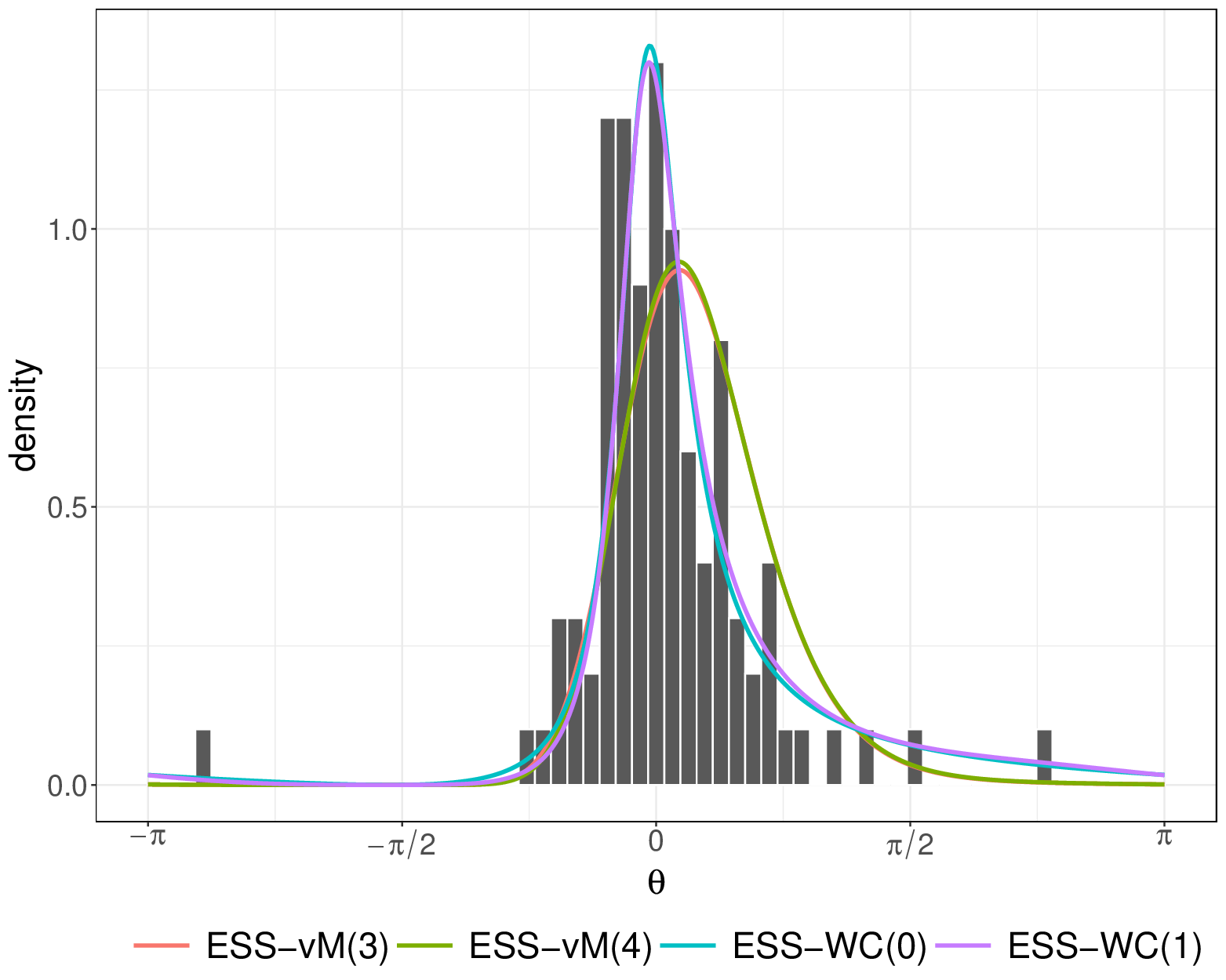}
		\caption{Histogram of dataset 2 with estimated density functions}
		\label{hist_fig2}
\end{figure}

Tables~\ref{dataset1} and \ref{dataset2} summarizes the estimated parameters with the AIC and TIC values for ESS-vM and ESS-WC distributions with $m\in \{0,1,2,3,4\}$. For Datasets 1 and 2, the AIC tends to select larger $m$, whereas the TIC selects smaller $m$, as can be confirmed from the simulation study. The density functions selected using AIC and TIC are shown in Figures \ref{hist_fig1} and \ref{hist_fig2}. According to these figures, using ESS-WC distributions has good fitting performances compared to ESS-vM distributions due to the highly concentrated data characteristics, and the fitted density curves seem to fit well for skewed data.

\begin{table}[htb]
\centering
\caption{Estimated parameters for Dataset 1}
\label{dataset1}
\begin{adjustbox}{max width=\textwidth}
\begin{tabular}{lrrrrrrrrrr}
  \hline \hline
  &  \multicolumn{5}{c}{ESS-vM}
    & \multicolumn{5}{c}{ESS-WC}\\
  $m$ & $\hat{\mu}$ & $\hat{\kappa}$ & $\hat{\lambda}$ & AIC & TIC
  & $\hat{\mu}$ & $\hat{\rho}$ & $\hat{\lambda}$ & AIC & TIC \\
  \hline
       0 & $-0.0170$ & 8.7164 & $-0.5721$ & $39.54$ & $\bm{41.77}$
         & $-0.0073$ & 0.8576 & $-0.6790$ & 35.34  & $\bm{35.46}$\\
       1 & $-0.0020$ & 8.4079 & $-0.5571$ & $39.43$ & 42.40 
         & $-0.0048$ & 0.8571 & $-0.5325$ & 35.18 & 35.86\\
       2 & 0.0082 & 8.1514 & $-0.5410$ & 39.33 & 42.79
         & $-0.0039$ & 0.8568 & $-0.4505$ & 35.13 & 36.03 \\
       3 & 0.0157 & 7.9431 & $-0.5246$ & 39.25& 43.05
         & $-0.0034$ & 0.8567 & $-0.3972$ &  35.10  & 36.11\\
       4 & 0.0214 & 7.7754 & $-0.5081$ & $\bm{39.19}$ & 43.23
         & $-0.0031$ & 0.8566 & $-0.3590$ & $\bm{35.09}$ & 36.17\\
       \hline 
   \end{tabular}
   \end{adjustbox}
\end{table}

\begin{table}[htb]
\centering
\caption{Estimated parameters for Dataset 2}
\label{dataset2}
\begin{adjustbox}{max width=\textwidth}
\begin{tabular}{lrrrrrrrrrr}
  \hline \hline
  &  \multicolumn{5}{c}{ESS-vM}
    & \multicolumn{5}{c}{ESS-WC}\\
  $m$ & $\hat{\mu}$ & $\hat{\kappa}$ & $\hat{\lambda}$ & AIC & TIC
  & $\hat{\mu}$ & $\hat{\rho}$ & $\hat{\lambda}$ & AIC & TIC \\
  \hline
     0 & $-0.0935$ & 3.8653 & 1.0000 & 143.92 &  143.90 
       & $-0.0688$ & 0.7834 & 1.0000 & 135.06 &  $\bm{133.19}$\\ 
     1 & $-0.1363$ & 3.4304 & 1.0000 & 139.23 &  138.36 
       & $-0.0900$ & 0.7753 & 0.9775 & $\bm{134.68}$ &  133.99 \\
     2 & $-0.1572$ & 3.2195 & 1.0000 & 136.78 &  135.61
       & $-0.0901$ & 0.7752 & 0.8013 & 134.93 &  134.37 \\
     3 & $-0.1693$ & 3.1002 & 1.0000 & 135.73 &  $\bm{134.46}$
       & $-0.0900$ & 0.7753 & 0.6926 & 135.04 &  134.52 \\
     4 & $-0.1751$ & 3.0449 & 0.9659 & $\bm{135.47}$ &  137.17
       & $-0.0899$ & 0.7753 & 0.6184 & 135.10 &  134.60\\
   \hline
\end{tabular}
   \end{adjustbox}
\end{table}

\section{Discussion}\label{Sec7}
We discuss the application of the proposed model to other statistical models and issues related to the choice of order $m$. 

Table \ref{tabBoundaryVM} in Section \ref{Sims} and Tables \ref{dataset1} and \ref{dataset2} in Section \ref{Sec6} show that as the order $m$ increases in the ESS distribution, an MLE of skewness parameter $\lambda$ is more likely to take values away from the boundary of parameter space. 
In general, it is known that when an MLE appears at the boundary of parameter space, the Hessian of minus the log-likelihood evaluated at the MLE may degenerate, which may affect the asymptotic variance of the MLE or make likelihood ratio tests unusable. Our results imply the possibility that such a method may also be applicable.

Table \ref{dataset2} also shows that the ESS distribution with some order $m$ larger than 0 can be fitted to a data distribution with a more skewed shape better than the usual sine-skewed circular distribution because it has a larger log-likelihood. 
From this advantage, the following finite mixture of the ESS circular distributions can represent stronger local asymmetric distributions than that in \cite{MSA20}: 
\begin{align*}
f_{\textrm{MESS}}(\theta )=\sum_{k=1}^{K}\pi_{k}f_{\textrm{ESS}}^{(m)}(\theta ;\mu_{k} ,\bm{\rho}_{k},\lambda_{k} ),
\end{align*}
where $\pi_{k}\geq 0$ $(k=1,\ldots ,K)$ are mixing proportions with $\sum_{k=1}^{K}\pi_{k}=1$, and $f_{\textrm{ESS}}^{(m)}(\theta ;\mu_{k} ,\bm{\rho}_{k},\lambda_{k} )$ is the ESS circular density \eqref{ESScircular} with parameters being $\mu_{k} ,\bm{\rho}_{k}$ and $\lambda_{k}$. 

In addition, because the skewing function $G_m(x)$ is a function of odd powers of $x$ such that $x$, $x^3$, $x^5$,..., by the same argument as in \cite{AL17}, we can give new cylindrical distributions combining the ESS-vM distribution in a circular part with the Weibull distribution in a linear part. The function $G_m(x)$ would also be used as skewing functions in skew-rotationally-symmetric distributions on the sphere proposed in \cite{LV17a}.

Next, we discuss the unimodality of the ESS circular distribution. As mentioned in Section \ref{Sec2}, the ESS-vM distribution of order $m$ $(m=0,1,2)$ does not satisfy the unimodality. Moreover, at this time, it is not known whether the ESS-WC distribution of order $m$ $(m\geq 1)$ is unimodal or not for all parameters. However, it is numerically confirmed that the ESS-WC density is unimodal in many cases of parameters.

Finally, we discuss the choice of order $m$. Although we introduced selection criteria based on AIC and TIC to select $m$, as shown by the simulations in Section \ref{Sims}, these criteria have difficulty in estimating the true $m$. As for TIC, when the estimated value of $\lambda$ is $1$ or $-1$, that is, when it does not satisfy the condition of taking the value of an interior point in a parameter space, the validity of its penalty term in TIC becomes problematic. Hence, it would be worthwhile to investigate alternative estimation methods for $m$, such as a Bayesian approach based on a marginal likelihood or a cross-validation method.

%\section*{Conflict of interest}
%The authors have no conflict of interest to declare. 

%\section*{Funding information}
%Yoichi Miyata was supported in part by JSPS KAKENHI Grant Number 19K11863. Takayuki Shiohama was supported in part by JSPS KAKENHI Grant Number 22K11944 and Nanzan University of Pache Research Subsidy I-A-2 for the 2022 academic year. Toshihiro Abe was supported in part by JSPS KAKENHI Grant Numbers 19K11869 and 19KK0287. 

\bibliographystyle{apalike} 
\bibliography{MSA2024paper}

\begin{thebibliography}{}

\bibitem[Abe et~al., 2022]{AISM21}
Abe, T., Imoto, T., Shiohama, T., and Miyata, Y. (2022).
\newblock On some flexible models for circular, toroidal, and cylindrical data.
\newblock In {\em Directional Statistics for Innovative Applications, A
  Bicentennial Tribute to Florence Nightingale}, pages 229--244. Springer.

\bibitem[Abe and Ley, 2017]{AL17}
Abe, T. and Ley, C. (2017).
\newblock A tractable, parsimonious and flexible model for cylindrical data,
  with applications.
\newblock {\em Econom. Stat.}, 4:91--104.

\bibitem[Abe and Pewsey, 2011]{AP11}
Abe, T. and Pewsey, A. (2011).
\newblock Sine-skewed circular distributions.
\newblock {\em Statist. Papers}, 52(3):683--707.

\bibitem[Abramowitz and Stegun, 1972]{AS72}
Abramowitz, M. and Stegun, I.~A. (1972).
\newblock {\em Handbook of Mathematical Functions with Formulas, Graphs, and
  Mathematical Tables}.
\newblock Dover, New York.

\bibitem[Balakrishnan, 2002]{Ba02}
Balakrishnan, N. (2002).
\newblock Discussion on "{S}kewmultivariatemodels related to hidden truncation
  and/orselective reporting" by {B}.{C}. arnold and {R}.{J}. beaver.
\newblock {\em TEST}, 11:37--39.

\bibitem[Batschelet, 1981]{batschelet1981circular}
Batschelet, E. (1981).
\newblock {\em Circular Statistics in Biology}.
\newblock Academic Press, New York.

\bibitem[Bekker et~al., 2022]{BRAL22}
Bekker, A., Nakhaei~Rad, N., Arashi, M., and Ley, C. (2022).
\newblock Generalized skew-symmetric circular and toroidal distributions.
\newblock In {\em Directional Statistics for Innovative Applications: A
  Bicentennial Tribute to Florence Nightingale}, pages 161--186. Springer.

\bibitem[Fisher, 1993]{Fi93}
Fisher, N.~I. (1993).
\newblock {\em Statistical Analysis of Circular Data}.
\newblock Cambridge University Press.

\bibitem[Jammalamadaka and Kozubowski, 2004]{JK04}
Jammalamadaka, S.~R. and Kozubowski, T.~J. (2004).
\newblock New families of wrapped distributions for modeling skew circular
  data.
\newblock {\em Commun. Stat.--Theory and Methods}, 33(9):2059--2074.

\bibitem[Kato and Jones, 2010]{KJ10}
Kato, S. and Jones, M.~C. (2010).
\newblock A family of distributions on the circle with links to, and
  applications arising from, {M}\"{o}bius transformation.
\newblock {\em J. Amer. Statist. Assoc.}, 105(489):249--262.

\bibitem[Kim and SenGupta, 2013]{KS13}
Kim, S. and SenGupta, A. (2013).
\newblock A three-parameter generalized von {M}ises distribution.
\newblock {\em Statist. Papers}, 54(3):685--693.

\bibitem[Konishi and Kitagawa, 2008]{konishi2008information}
Konishi, S. and Kitagawa, G. (2008).
\newblock {\em Information Criteria and Statistical Modeling}.
\newblock Springer.

\bibitem[Ley et~al., 2021]{LBC21}
Ley, C., Babi{\'c}, S., and Craens, D. (2021).
\newblock Flexible models for complex data with applications.
\newblock {\em Annual Review of Statistics and Its Application}, 8:369--391.

\bibitem[Ley and Verdebout, 2017]{LV17a}
Ley, C. and Verdebout, T. (2017).
\newblock Skew-rotationally-symmetric distributions and related efficient
  inferential procedures.
\newblock {\em J. Multivariate Anal.}, 159:67--81.

\bibitem[Mardia and Jupp, 2000]{MJ09}
Mardia, K.~V. and Jupp, P.~E. (2000).
\newblock {\em Directional Statistics}.
\newblock John Wiley \& Sons, Ltd., Chichester.

\bibitem[Miyata et~al., 2020]{MSA20}
Miyata, Y., Shiohama, T., and Abe, T. (2020).
\newblock Estimation of finite mixture models of skew-symmetric circular
  distributions.
\newblock {\em Metrika}, 83(8):895--922.

\bibitem[Miyata et~al., 2022]{MSA19}
Miyata, Y., Shiohama, T., and Abe, T. (2022).
\newblock Identifiability of asymmetric circular and cylindrical distributions.
\newblock {\em Sankhya A. to appear}.

\bibitem[Pewsey, 2002]{pewsey2002testing}
Pewsey, A. (2002).
\newblock Testing circular symmetry.
\newblock {\em Can. J. Stat.}, 30(4):591--600.

\bibitem[Umbach and Jammalamadaka, 2009]{UJ09}
Umbach, D. and Jammalamadaka, S.~R. (2009).
\newblock Building asymmetry into circular distributions.
\newblock {\em Statist. Probab. Lett.}, 79(5):659--663.

\bibitem[van~der Vaart, 2000]{VV20}
van~der Vaart, A.~W. (2000).
\newblock {\em Asymptotic Statistics}.
\newblock Cambridge University Press.

\bibitem[Yfantis and Borgman, 1982]{yfantis1982extension}
Yfantis, E. and Borgman, L. (1982).
\newblock An extension of the von {M}ises distribution.
\newblock {\em Commun. Stat.--Theory and Methods}, 11(15):1695--1706.

\end{thebibliography}

\appendix
\section{Proofs of the results}\label{appendix}

\subsection{Proof of Proposition \ref{prop_identi2}} 
Due to the complexity of proving identifiability for ESS-WC distributions of general order $m$, we initially establish the identifiability specifically for ESS-WC distributions of order $2$. Subsequently, we extend our proof to encompass the identifiability of ESS-WC distributions for any general order $m$. 

Using the cosine and sine moments (\ref{ESSWC2cosine}) and (\ref{ESSWC2sine}), the squared $p$th mean resultant length $\rho_{\textrm{MRL},p}^{2}$ is is expressed as
\begin{align*}
\rho_{\textrm{MRL},p}^{2}&=\biggl\{ \frac{15}{128}c_{2,1}(\lambda )\left( \rho^{|p+1|}-\rho^{|p-1|}\right) +\frac{5}{256}c_{2,2}(\lambda )\left( \rho^{|p+3|}-\rho^{|p-3|}\right)\notag \\
 &-\frac{3}{256}\lambda^{5}\left( \rho^{|p+5|}-\rho^{|p-5|}\right) \biggr\}^{2} +\rho^{2|p|}.
\end{align*}

Here, we divide the condition $\bm{\eta}_{1}\ne\bm{\eta}_{2}$ into three cases.
\begin{description}
\setlength{\parskip}{0cm} 
\setlength{\itemsep}{0cm} 
\item[Case 1] $\rho_{1}\ne \rho_{2}$
\item[Case 2] $\rho_{1}=\rho_{2}$ and $\mu_{1}\ne \mu_{2}$
\item[Case 3] $\rho_{1}=\rho_{2}$, $\mu_{1}= \mu_{2}$, and $\lambda_{1}\ne \lambda_{2}$
\end{description}

In Case 1, we consider the case $\rho_{1}<\rho_{2}$ and $p\geq 5$. 
Then, the ratio of the squared $p$th MRL $\rho_{\textrm{MRL},p}(\bm{\eta}_{1})^{2}$ to $\rho_{\textrm{MRL},p}(\bm{\eta}_{2})^{2}$ and taking the limit as $p\to\infty$ yields
\begin{align*}
\frac{\rho_{\textrm{MRL},p}(\bm{\eta}_{1})^{2}}{\rho_{\textrm{MRL},p}(\bm{\eta}_{2})^{2}}&=\frac{M_{1}(\rho_{1},\lambda_{1})}{M_{1}(\rho_{2},\lambda_{2})}\left( \frac{\rho_{1}}{\rho_{2}}\right)^{2p}\to 0\quad (p\to\infty ),
\end{align*}
where $M_{1}(\rho ,\lambda )=(15/128)c_{2,1}(\lambda )( \rho-\rho^{-1}) +(5/256)c_{2,2}(\lambda )( \rho^{3}-\rho^{-3})-(3/256)\lambda^{5}( \rho^{5}-\rho^{-5})$ is a function of $(\rho ,\lambda )$ and does not depend on $p$.
In the case $\rho_{1}>\rho_{2}$, we prove it similarly.

Next, we consider Case 2 with $\rho_{1}=\rho_{2}$ and $\mu_{1}\ne \mu_{2}$.
Let $\Psi^{*} (\bm{\eta})$ be the characteristic function of the distribution and let $p\geq 5$. Then, we have
\begin{align*}
\Psi^{*} (\bm{\eta})&=\int_{\mu}^{2\pi +\mu}\exp(ip\theta )f_{\textrm{ESSWC}}^{(2)}(\theta ;\bm{\eta})d\theta \\
 &=\exp(ip\mu )\left( \alpha_{p}(\rho )+i\beta_{p}(\rho ,\lambda )\right) ,
\end{align*}
where $\alpha_{p}(\rho )$ is the $p$th cosine moment under $\mu=0$, and $\beta_{p}(\rho ,\lambda )$ is the $p$th sine moment under $\mu=0$.
We prove $\Psi^{*} (\bm{\eta}_{1})\ne \Psi^{*} (\bm{\eta}_{2})$ by contradiction. Hence, assume that $\Psi^{*} (\bm{\eta}_{1})= \Psi^{*} (\bm{\eta}_{2})$. 
Then, it follows that 
\begin{align}
\exp(ip\mu_{1} )\left( \alpha_{p}(\rho ,\lambda_{1})+i\beta_{p}(\rho ,\lambda_{1} )\right) &=\exp(ip\mu_{2} )\left( \alpha_{p}(\rho ,\lambda_{2} )+i\beta_{p}(\rho ,\lambda_{2} )\right) , \label{Pr:A2}
\end{align}
where $\mu_{1}\ne \mu_{2}$, $\mu_{1}-\mu_{2}\in (-2\pi ,2\pi )\backslash \{ 0\}$. 
When $\mu_{1}-\mu_{2}\in (0,2\pi )$,  Equation \eqref{Pr:A2} becomes
\begin{align}
\exp\{ ip(\mu_{1}-\mu_{2} )\}\rho^{p}\left( 1+M_{1}(\rho ,\lambda_{1})i\right) &=\rho^{p}\left( 1+M_{1}(\rho ,\lambda_{2})i\right) .
\end{align}
Dividing the equation by $\rho^{p}$, we have
\begin{align}
\frac{1}{n}\sum_{p=5}^{n+4}\exp\{ ip(\mu_{1}-\mu_{2} )\} \left( 1+M_{1}(\rho ,\lambda_{1})i\right) &= 1+M_{1}(\rho ,\lambda_{2})i .\label{Pr:A3}
\end{align}
As $\mu_{1}-\mu_{2} \ne 0$, we have
\begin{align}
\frac{1}{n}\sum_{p=5}^{n+4}\exp\{ ip(\mu_{1}-\mu_{2} )\} &=\exp\{ i 5(\mu_{1}-\mu_{2})\} \frac{1}{n}\frac{1-\exp\{ i(n-1)(\mu_{1}-\mu_{2})\}}{1-\exp\{ i(\mu_{1}-\mu_{2})\}} \notag \\
 &\to 0\qquad (n\to \infty).
\end{align}
Hence, taking the limit as $n\to \infty$ in Equation \eqref{Pr:A3}, we have
\begin{align*}
0=1+M_{1}(\rho ,\lambda_{2})i,
\end{align*}
which is a contradiction. 

Finally, we consider Case 3 with $\rho_{1}=\rho_{2}$, $\mu_{1}= \mu_{2}$, and $\lambda_{1}\ne \lambda_{2}$. 
For simplicity, let $f_{\textrm{WC}}(\theta ;\mu , \rho )=(1-\rho^{2})/[(2\pi )\{ 1+\rho^{2}-2\rho \cos (\theta -\mu )\} ]$ be the density of the WC distribution.
We prove $f_{\textrm{ESSWC}}^{(2)}(\theta ;\bm{\eta}_{1})\ne f_{\textrm{ESSWC}}^{(2)}(\theta ;\bm{\eta}_{2})$ by contradiction. 
If we assume $f_{\textrm{ESSWC}}^{(2)}(\theta ;\bm{\eta}_{1})= f_{\textrm{ESSWC}}^{(2)}(\theta ;\bm{\eta}_{2})$, we have $G_{m}(\lambda_{1} \sin (\theta -\mu ))=G_{m}(\lambda_{2} \sin (\theta -\mu ))$. 
As $G_{m}(x)$ is a strictly monotonically increasing function of $x$, we have $\lambda_{1} \sin (\theta -\mu )=\lambda_{2} \sin (\theta -\mu )$ for any $\theta$. Hence, we have $\lambda_{1}=\lambda_{2}$, which contradicts the assumption. 
Therefore, the identifiability for the case $m=2$ is proved.

Subsequently, we present the $p$th mean resultant length when order $m$ is general. It follows from de Moivre's theorem that
\begin{align}
\cos (n\theta )+i\sin (n\theta )=\left( \cos\theta +i\sin\theta\right)^{n} 
 &=\sum_{k=0}^{n}\binom{n}{k}\cos^{n}\theta (i\sin \theta )^{k}.\label{Pr:A4}
\end{align}
Here, we consider only the case where $n$ is an odd number, and hence let $n=2\ell +1$.
Focusing on the imaginary part of the above equation in Equation \eqref{Pr:A4}, we have
\begin{align}
\sin (n\theta )&=\binom{n}{1}\cos^{n-1}\theta \sin\theta +\binom{n}{3}\cos^{n-3}\theta \sin^{3}\theta (i^{2})\notag \\
  &+\cdots +\binom{n}{n-2}\cos^{2}\theta \sin^{n-2}\theta (i^{n-2})+\sin^{n}\theta (i^{n-1}).
\end{align}
Note that each of $i^{n-3}$ and $i^{n-1}$ has a real value $-1$ or $1$ since $n$ is odd. 
Substituting $n=2\ell +1$ and using $\cos^{2} \theta =1-\sin^{2}\theta$, we have
\begin{align}
\sin (n\theta )=&a_{n,2\ell +1} \sin^{2\ell +1}\theta +a_{n,2\ell -1}\sin^{2\ell -1}\theta +a_{n,2\ell -3}\sin^{2\ell -3}\theta +\cdots +a_{1}\sin\theta ,\notag
\end{align}
where $a_{n,j}$ $(j=1,3,\ldots ,2\ell -3 ,2\ell -1 )$ are constants, and
\begin{align}
a_{n,2\ell +1}&=(-1)^{\ell}\binom{n}{1}-(-1)^{\ell -1}\binom{n}{3}+(-1)^{\ell -2}\binom{n}{5}+\cdots -(-1)^{\ell -1}\binom{n}{n-2}+(-1)^{\ell}\notag \\
 &=(-1)^{\ell}2^{n-1}. \notag
\end{align}
Hence, we have
\begin{align}
\sin (n\theta )&=a_{n,n}\sin^{n}\theta +a_{n,n-2}\sin^{n-2}\theta +\cdots +a_{n,3}\sin^{3}\theta +a_{n,1}\sin\theta .
\end{align}
Thus, we have
\begin{align}
\begin{pmatrix}
\sin (n\theta ) \\
\sin((n-2)\theta ) \\
\vdots \\
\sin 3\theta \\
\sin \theta
\end{pmatrix}
&=
\begin{pmatrix}
  a_{n,n}          & a_{n,n-2} & \dots  & a_{n,1} \\
                  & a_{n-2,n-2} & \dots  & a_{n-2,1} \\
                  &        & \ddots & \vdots \\
  \text{\huge{0}} &        &        & a_{1,1}
\end{pmatrix}
\begin{pmatrix}
\sin^{n}\theta \\
\sin^{n-2}\theta  \\
\vdots \\
\sin^{3}\theta \\
\sin \theta
\end{pmatrix}.
\end{align}
For simplicity, we let the upper triangular matrix on the right-hand side of the above equation be $\bm{A}$. 
Because $a_{j,j}\ne 0$ $(j=1,2,\ldots ,n)$, we have $\textrm{det}(\bm{A})=a_{1,1}a_{3,3}\cdots a_{n-2,n-2}a_{n,n}\ne 0$. 
Hence, 
\begin{align}
\begin{pmatrix}
\sin^{n}\theta \\
\sin^{n-2}\theta  \\
\vdots \\
\sin^{3}\theta \\
\sin \theta
\end{pmatrix}
&=\bm{A}^{-1}
\begin{pmatrix}
\sin (n\theta ) \\
\sin((n-2)\theta ) \\
\vdots \\
\sin 3\theta \\
\sin \theta
\end{pmatrix}
\end{align}
From this result, $\sin^{n}\theta$ is expressed as a linear combination of $\sin (n\theta )$, $\sin ((n-2)\theta ),\ldots ,\sin (3\theta )$, and $\sin \theta$. Hence, if $n$ is odd, we can write 
\begin{align}
\sin^{n}\theta &=\sum_{k=0}^{(n-1)/2}c_{n,n-2k}\sin ((n-2k)\theta) \label{c_nk}
\end{align}
for some constants $c_{n,n-2k}$ $(k=0,\ldots ,(n-1)/2)$. Note that $c_{n,n}=1/a_{n,n}\ne 0$ using the standard matrix algebra.

For simplicity, we let $n(\ell )=2\ell +1$ and let the base density $f_{0}(\theta ;\rho )$ be the wrapped Cauchy density. Then, the $p$th sine moment \eqref{beta_p} with $\mu =0$ and order $m$ can be expressed as follows:
\begin{align}
\beta_{p}&=2C_{m}\sum_{\ell=0}^{m}\binom{m}{\ell}\frac{(-1)^{\ell}}{2\ell+1} \lambda^{2\ell +1}\int_{-\pi}^{\pi}\sum_{k=0}^{\ell}c_{n(\ell ),n(\ell )-2k}\sin\left( (n(\ell ) -2k)\theta\right)\sin (p\theta )f_{0}(\theta ;\rho )d\theta \notag \\
 &=2C_{m}\sum_{\ell=0}^{m}\binom{m}{\ell}\frac{(-1)^{\ell}}{2\ell+1} \lambda^{2\ell +1}\sum_{k=0}^{\ell}c_{n(\ell ),n(\ell )-2k}\notag \\
 &\times \int_{-\pi}^{\pi}\left( -\frac{1}{2}\right) \left\{ \cos ((n(\ell )-2k+p)\theta ) -\cos ((n(\ell )-2k-p)\theta )\right\} f_{0}(\theta ;\rho )d\theta \notag \\
 &=C_{m}\sum_{\ell=0}^{m}\binom{m}{\ell}\frac{(-1)^{\ell +1}}{2\ell+1} \lambda^{2\ell +1} \sum_{k=0}^{\ell}c_{n(\ell ),n(\ell )-2k}\left\{ \rho^{|n(\ell )-2k+p|}-\rho^{|n(\ell )-2k-p|}\right\} , \label{proof:sine_moment1}
\end{align}
where $c_{n(\ell),n(\ell )}\ne 0$ for $\ell \in \{0,\ldots ,m\}$. 
If $p$ is a sufficiently large positive integer, the expression \eqref{proof:sine_moment1} is written as $B_{m}\rho^{p}$ where $B_{m}$ is a constant independent of $p$. From this result and Equation \eqref{alpha_p}, the $p$th cosine and sine moments are $\alpha_{p,\mu}=-\sin (p\mu )B_{m}\rho^{p}+\cos (p\mu )\rho^{p}$ and $\beta_{p,\mu}=\cos (p\mu )B_{m}+\sin (p\mu )\rho^{p}$.
Hence, the $p$th mean result length is expressed as 
\begin{align*}
\rho_{\textrm{MRL},p}=\sqrt{B_{m}^{2}+1}\rho^{p} .
\end{align*}
Using this equation and following the same manner described above, the identifiability of the family $\mathcal{F}_{\textrm{WC}}^{(m)}$ holds for the general order $m$. \hfill $\square$

\subsection{Proof of Proposition \ref{prop_identi1}}
We present a lemma to prove Proposition \ref{prop_identi1}.

\begin{lemma}\label{Proof:lemA1}
\begin{enumerate}
\item For any $\tau \in\mathbb{N}$, $I_{p+\tau }(\kappa )/I_{p}(\kappa )\to 0$ as $p\to\infty$.
\item For any $\tau \in\mathbb{N}$,
\begin{align*}
\frac{I_{p+\tau}(\kappa )-I_{p-\tau}(\kappa )}{I_{p}(\kappa )p^{\tau}}\to -\frac{2^\tau}{\kappa^\tau}
\end{align*}
as $p\to \infty$.
\end{enumerate}
\end{lemma}
{\sc Proof}. From equation (9.6.19) on page 376 of \cite{AS72}, it follows that
\begin{align*}
I_{p+1}(\kappa )&=\sum_{r=0}^{\infty}\frac{1}{(p+r+1)! r!}\left( \frac{\kappa}{2}\right)^{2r+p+1} \\
 &=\sum_{r=0}^{\infty}\frac{\kappa /2}{p+r+1}\frac{1}{(p+r)! r!}\left( \frac{\kappa}{2}\right)^{2r+p}\\
 &\leq \frac{\kappa /2}{p+1}\sum_{r=0}^{\infty}\frac{1}{(p+r)! r!}\left( \frac{\kappa}{2}\right)^{2r+p}\\
 &=\frac{\kappa}{2(p+1)}I_{p}(\kappa ).
\end{align*}
Hence, we have
\begin{align*}
0\leq \frac{I_{p+1}(\kappa )}{I_{p}(\kappa )}\leq \frac{\kappa}{2(p+1)},
\end{align*}
which leads to the first statement.

It follows from equation (9.6.26) on page 376 of \cite{AS72} or equations (A.8) and (A.9) on page 350 of \cite{MJ09} that for any $\tau\in\mathbb{N}$,
\begin{align}
I_{p-\tau}(\kappa )&=\frac{2(p-\tau+1)}{\kappa}I_{p-\tau +1}(\kappa)+I_{p-\tau+2}(\kappa). \label{eqA.1}
\end{align}
In order to make the outline of the proof easier to understand, we consider the case $\tau=3$. 

Using Equation \eqref{eqA.1} repeatedly, we have
\begin{align*}
I_{p-3}(\kappa)&=\frac{2(p-2)}{\kappa}I_{p-2}(\kappa)+I_{p-1}(\kappa) \\
               &=\frac{2(p-2)}{\kappa}\left\{ \frac{2(p-1)}{\kappa}I_{p-1}(\kappa)+I_{p}(\kappa )\right\} +I_{p-1}(\kappa) \\
    &=\frac{2(p-2)}{\kappa}\left\{ \frac{2(p-1)}{\kappa}\left\{ \frac{2p}{\kappa}I_{p}(\kappa)+I_{p+1}(\kappa )\right\} +I_{p}(\kappa )\right\} +\frac{2p}{\kappa}I_{p}(\kappa)+I_{p+1}(\kappa ) .
\end{align*}
Thus, using the first statement of Lemma \ref{Proof:lemA1}, we have
\begin{align*}
\frac{I_{p-3}(\kappa)}{I_{p}(\kappa)p^{3}}\to \frac{2^{3}}{\kappa^3} \qquad (p\to \infty ).
\end{align*}
Hence, using this result again, we have 
\begin{align*}
\frac{I_{p+3}(\kappa )-I_{p-3}(\kappa )}{I_{p}(\kappa )p^{3}}\to -\frac{2^3}{\kappa^3}, \qquad (p\to\infty).
\end{align*}

Now, using induction, we prove the general case. 
Assume that for fixed $\tau \in\mathbb{N}$,
\begin{align*}
\frac{I_{p-(\tau -1)}(\kappa )}{I_{p}(\kappa )p^{\tau -1}}\to \frac{2^{\tau -1}}{\kappa^{\tau-1}}, \frac{I_{p-(\tau -2)}(\kappa )}{I_{p}(\kappa )p^{\tau -1}}\to \frac{2^{\tau -2}}{\kappa^{\tau-2}},\cdots ,\frac{I_{p-1}(\kappa )}{I_{p}(\kappa )p}\to \frac{2}{\kappa},
\end{align*}
as $p\to\infty$. 
Then, Equation \eqref{eqA.1} repeatedly, we have

\begin{align*}
I_{p-\tau}(\kappa)=\frac{2(p-\tau +1)}{\kappa}\left\{ \frac{2(p-\tau +2)}{\kappa}\left\{ \cdots \left\{ \frac{2p}{\kappa}I_{p}(\kappa)+I_{p+1}(\kappa)\right\}\cdots \right\} +I_{p-\tau+3}(\kappa) \right\} +I_{p-\tau+2}(\kappa) .
\end{align*}
Dividing this equation by $I_{p}(\kappa )p^{\tau}$ and taking the limit, we have\begin{align*}
\frac{I_{p-\tau}(\kappa)}{I_{p}(\kappa)p^{\tau}}\to \frac{2^{\tau}}{\kappa^{\tau}}\qquad (p\to \infty),
\end{align*}
which completes the proof.\hfill $\square$

After proving the identifiability of the family of ESS-vM distributions of order $2$, we give a proof for the general order $m$ by the same argument. 

For simplicity, let $\tilde{H}_{s}^{p}(\kappa):=\left\{ I_{p+s}(\kappa )-I_{p-s}(\kappa )\right\} /I_{p}(\kappa )$ and let 
\begin{align}
S_{1,p}(\bm{\eta})&=\frac{15}{128}c_{2,1}(\lambda)\left( \frac{I_{p+1}(\kappa )-I_{p-1}(\kappa)}{I_{p}(\kappa)}\right) +\frac{5}{256}c_{2,2}(\lambda)\left( \frac{I_{p+3}(\kappa )-I_{p-3}(\kappa)}{I_{p}(\kappa)}\right) \notag\\
                &-\frac{3}{256}\lambda^{5}\left( \frac{I_{p+5}(\kappa )-I_{p-5}(\kappa)}{I_{p}(\kappa)}\right) \notag \\
 &=\frac{15}{128}c_{2,1}(\lambda)\tilde{H}_{1}^{p}(\kappa) +\frac{5}{256}c_{2,2}(\lambda)\tilde{H}_{3}^{p}(\kappa)-\frac{3}{256}\lambda^{5}\tilde{H}_{5}^{p}(\kappa) ,
  \notag
\end{align}
where $c_{2,j}(\lambda)$ $(j=1,2)$ are defined in Proposition \ref{prop_identi1}. Then, the $p$th cosine moments with $\mu =0$ are rewritten as 
\begin{align}
\beta_{p} &=S_{1,p}(\bm{\eta})\frac{I_{p}(\kappa )}{I_{0}(\kappa )}.
\end{align}
Let $\bm{\eta}_{i}=(\mu_{i},\kappa_{i},\lambda_{i})$ $(i=1,2)$. 
First, we divide the condition $\bm{\eta}_{1}\ne\bm{\eta}_{2}$ into three cases.
\begin{description}
\setlength{\parskip}{0cm} 
\setlength{\itemsep}{0cm} 
\item[Case 1] $\kappa_{1}\ne \kappa_{2}$
\item[Case 2] $\kappa_{1}=\kappa_{2}$ and $\lambda_{1}\ne \lambda_{2}$
\item[Case 3] $\kappa_{1}=\kappa_{2}$, $\lambda_{1}= \lambda_{2}$, and $\mu_{1}\ne \mu_{2}$
\end{description}
In Case 1, we consider the case $\kappa_{1}<\kappa_{2}$. Let $\rho_{\textrm{MRL},p}(\bm{\eta}_{i})$ be the $p$th mean resultant length for parameter vector $\bm{\eta}_{i}$. 
It follows from Lemma \ref{Proof:lemA1} that $\tilde{H}_{s}^{p}(\kappa)=O\left( p^{s}\right)$ $(s=1,3,5)$.
Hence, $S_{1,p}(\bm{\eta}_{1})^{2}$ increases in polynomial order of $p$. 

On the other hand, it follows from Lemma 9 of \citet{MSA19} that
\begin{align}
\frac{I_{p}(\kappa_{1})}{I_{p}(\kappa_{2})}=O\left( \left( \frac{\kappa_{1}}{\kappa_{2}}\right)^{p}\right),
\end{align}
which implies that $(I_{p}(\kappa_{1})/I_{p}(\kappa_{2}))^2$ decreases to zero in exponential order if $\kappa_{1}<\kappa_{2}$.
Therefore, we have
\begin{align}
\frac{\rho_{\textrm{MRL},p}(\bm{\eta}_{1})^{2}}{\rho_{\textrm{MRL},p}(\bm{\eta}_{2})^{2}}&=O(p^{10})O\left( \left( \frac{\kappa_{1}}{\kappa_{2}}\right)^{2p}\right) \to 0
\end{align}
as $p\to \infty$. The case with $\kappa_{2}<\kappa_{1}$ is also proved similarly. 

Second, we consider the case with $\kappa_{1}=\kappa_{2}=\kappa$ and $\lambda_{1}\ne \lambda_{2}$. 
We write the characteristic function of the ESS-vM distribution by
\begin{align*}
\Psi (\bm{\eta})&=\exp (ip\mu )\left( \alpha_{p}(\kappa )+i\beta_{p}(\kappa ,\lambda )\right) ,
\end{align*}
where $\alpha_{p}(\kappa )$ is the same as expression \eqref{ESSvM2cos} and $\beta_{p}(\kappa ,\lambda)$ is the same as expression \eqref{ESSvM2sin}. 
If $\lambda_{1}=0$, it is obvious that $\Psi (\bm{\eta}_{1})\ne \Psi (\bm{\eta}_{2})$ because $\beta_{p}(\kappa ,0)=0$. Therefore, without loss of generality, we assume $\lambda_{1}\ne 0$.
Now, we prove $\Psi (\bm{\eta}_{1})\ne \Psi (\bm{\eta}_{2})$ by contradiction.
To do that, we assume $\Psi (\bm{\eta}_{1})=\Psi (\bm{\eta}_{2})$.
Then, we have
\begin{align}
\exp (ip\mu_{1} )\left( \alpha_{p}(\kappa )+i\beta_{p}(\kappa ,\lambda_{1} )\right) &=\exp (ip\mu_{2} )\left( \alpha_{p}(\kappa )+i\beta_{p}(\kappa ,\lambda_{2} )\right) ,
\end{align}
where $-2\pi <\mu_{1}-\mu_{2}<2\pi$. 
%We consider the case $0\leq \mu_{1}-\mu_{2}<2\pi$. 
Then, 
\begin{align}
&\exp (ip(\mu_{1}-\mu_{2}))=\frac{ \alpha_{p}(\kappa )+i\beta_{p}(\kappa ,\lambda_{2} )}{\alpha_{p}(\kappa )+i\beta_{p}(\kappa ,\lambda_{1} )}\label{Proof:ESSvM_eq1} \\
 &=\frac{\alpha_{p}(\kappa )^{2}-\beta_{p}(\kappa ,\lambda_{2})\beta_{p}(\kappa ,\lambda_{1} )+i\left\{ \alpha_{p}(\kappa)\beta_{p}(\kappa ,\lambda_{2})-\alpha_{p}(\kappa)\beta_{p}(\kappa ,\lambda_{1})\right\}}{\rho_{\textrm{MRL},p}(\bm{\eta}_{1})^{2}}. \label{Proof:ESSvM_eq2}
\end{align}
The real part of expression \eqref{Proof:ESSvM_eq2} is
\begin{align}
&\biggl[ 1-\left\{ \frac{15}{128}c_{2,1}(\lambda_{2})\tilde{H}_{1}^{p}(\kappa ) +\frac{5}{256}c_{2,2}(\lambda_{2})\tilde{H}_{3}^{p}(\kappa ) -\frac{3}{256}\lambda_{2}^{5}\tilde{H}_{5}^{p}(\kappa ) \right\} \notag \\
\times &\left\{ \frac{15}{128}c_{2,1}(\lambda_{1})\tilde{H}_{1}^{p}(\kappa ) +\frac{5}{256}c_{2,2}(\lambda_{1})\tilde{H}_{3}^{p}(\kappa ) -\frac{3}{256}\lambda_{1}^{5}\tilde{H}_{5}^{p}(\kappa ) \right\} \biggr] \notag \\
&\times \frac{1}{1+\left\{ \frac{15}{128}c_{2,1}(\lambda_{1})\tilde{H}_{1}^{p}(\kappa ) +\frac{5}{256}c_{2,2}(\lambda_{1})\tilde{H}_{3}^{p}(\kappa ) -\frac{3}{256}\lambda_{1}^{5}\tilde{H}_{5}^{p}(\kappa ) \right\}^{2} }.
\end{align}
Because $\tilde{H}_{1}^{p}(\kappa )=O(p)$, $\tilde{H}_{3}^{p}(\kappa )=O(p^3)$ and $(1/p^{5})\tilde{H}_{5}^{p}(\kappa )\to -2^5 /\kappa^{5}$ as $p\to\infty$, we have
\begin{align}
\frac{\alpha_{p}(\kappa )^{2}-\beta_{p}(\kappa ,\lambda_{2})\beta_{p}(\kappa ,\lambda_{1} )}{\rho_{\textrm{MRL},p}(\bm{\eta}_{1})^{2}}&\to \frac{-(3/256)\lambda_{2}^{5}(32/\kappa^{5})\left( -(3/256)\lambda_{1}^{5}(32/\kappa^{5})\right)}{\left( -(3/256)\lambda_{2}^{5}(32/\kappa^{5})\right)^{2}} \notag \\
&=\frac{\lambda_{2}^{5}}{\lambda_{1}^{5}}\ne 1 . \label{Proof:ESSvM_eq3}
\end{align}
Next, we look at the left side of Equation \eqref{Proof:ESSvM_eq1}, that is $\exp (ip(\mu_{1}-\mu_{2}))$. 
From \cite{MSA20}, there exists a subsequence $\{ p_{n}\}$ such that $p_{n}(\mu_{1}-\mu_{2})\to 0$ (mod $2\pi$) as $n\to\infty$.
Accordingly, we have 
\begin{align}
\exp (ip_{n}(\mu_{1}-\mu_{2})) \to 1 ,\qquad (n\to\infty ). \label{Proof:ESSvM_eq4}
\end{align}
As the limit \eqref{Proof:ESSvM_eq3} also holds for the subsequence $\{ p_{n}\}$, this contradicts the result \eqref{Proof:ESSvM_eq4}.

Finally, we consider the case with $\kappa_{1}=\kappa_{2}=\kappa$, $\lambda_{1}= \lambda_{2}=\lambda$, and $\mu_{1}\ne \mu_{2}$. 
We prove $\Psi (\bm{\eta}_{1})\ne \Psi (\bm{\eta}_{2})$ by contradiction.
Assume $\Psi (\bm{\eta}_{1})=\Psi (\bm{\eta}_{2})$. Then, we have
\begin{align*}
\exp \left( ip(\mu_{1}-\mu_{2})\right)=1 \qquad (p\in\mathbb{Z}).
\end{align*}
Hence,
\begin{align*}
\frac{1}{N}\sum_{p=0}^{N-1}\exp \left( ip(\mu_{1}-\mu_{2})\right)=1.
\end{align*}
Taking the limit as $N\to\infty$ leads to $0=1$, which leads to contradiction.

Next, we consider the case when the order $m$ is general.
To simplify notation, let $n(\ell )=2\ell +1$. 
Remind that $p$th cosine moment with $\mu =0$ is $\alpha_{p}=I_{p}(\kappa )/I_{0}(\kappa )$. $C_m$ is defined in Section 2 and $c_{n(\ell ),n(\ell )-2k}$ is given in Equation \eqref{c_nk}. 
The $p$th sine moment with $\mu =0$ can be expressed as
\begin{align}
\beta_{p}&=2C_{m}\sum_{\ell=0}^{m}\binom{m}{\ell}\frac{(-1)^{\ell}}{2\ell+1} \lambda^{2\ell +1}\int_{-\pi}^{\pi}\sum_{k=0}^{\ell}c_{n(\ell ),n(\ell )-2k}\sin\left( (n(\ell ) -2k)\theta\right)\sin (p\theta )f_{0}(\theta ;\rho )d\theta \notag \\
 &=2C_{m}\sum_{\ell=0}^{m}\binom{m}{\ell}\frac{(-1)^{\ell}}{2\ell+1} \lambda^{2\ell +1}\sum_{k=0}^{\ell}c_{n(\ell ),n(\ell )-2k}\notag \\
 &\times \int_{-\pi}^{\pi}\left( -\frac{1}{2}\right) \left\{ \cos ((n(\ell )-2k+p)\theta ) -\cos ((n(\ell )-2k-p)\theta )\right\} f_{0}(\theta ;\rho )d\theta \notag \\
 &=C_{m}\sum_{\ell=0}^{m}\binom{m}{\ell}\frac{(-1)^{\ell +1}}{2\ell+1} \lambda^{2\ell +1} \sum_{k=0}^{\ell}c_{n(\ell ),n(\ell )-2k}\left\{ \frac{I_{n(\ell )-2k+p}(\kappa )-I_{n(\ell )-2k-p}(\kappa )}{I_{p}(\kappa)}\right\} \frac{I_{p}(\kappa )}{I_{0}(\kappa )} , \notag \\
 &=C_{m}\sum_{\ell=0}^{m}\binom{m}{\ell}\frac{(-1)^{\ell +1}}{2\ell+1} \lambda^{2\ell +1} \sum_{k=0}^{\ell}c_{n(\ell ),n(\ell )-2k}\tilde{H}^{p}_{n(\ell)-2k}(\kappa) \frac{I_{p}(\kappa )}{I_{0}(\kappa )} , \label{proof:sine_moment1} \\
 &=S_{1,p}(\kappa ,\lambda )\frac{I_{p}(\kappa )}{I_{0}(\kappa )},
\end{align}
where $\tilde{H}^{p}_{n(\ell)-2k}(\kappa)=(I_{n(\ell )-2k+p}(\kappa )-I_{n(\ell )-2k-p}(\kappa ))/I_{p}(\kappa)$ and $S_{1,p}(\kappa ,\lambda )$ is a part except for $I_{p}(\kappa )/I_{0}(\kappa )$ in Equation \eqref{proof:sine_moment1}. 
Notice taht $\tilde{H}^{p}_{n(\ell)-2k}(\kappa)=O\left( p^{n(\ell)-2k}\right)$ and has the largest order of $p$ when $\ell =m$ and $k=0$.

Let $\rho_{\textrm{MRL},p}(\bm{\eta})$ be the $p$th mean resultant length for parameter vector $\bm{\eta}$. Then, we have
\begin{align*}
\rho_{\textrm{MRL},p}(\bm{\eta})^2&=\alpha_{p}^{2}+\beta_{p}^{2} \\
 &=\left\{ 1+S_{1,p}(\kappa ,\lambda)^{2}\right\} \frac{I_{p}(\kappa )^{2}}{I_{0}(\kappa )^{2}}.
\end{align*}

Now, for different two parameter vectors $\bm{\eta}_{1}=(\mu_{1},\kappa_{1},\lambda_{1})^T$ and $\bm{\eta}_{2}=(\mu_{2},\kappa_{2},\lambda_{2})^T$, 
under the following cases, we show $f_{ESSvM}(\theta ;\bm{\eta}_{1})\ne f_{ESSvM}(\theta ;\bm{\eta}_{2})$.

\begin{description}
\setlength{\parskip}{0cm} 
\setlength{\itemsep}{0cm} 
\item[Case 1] $\kappa_{1}\ne \kappa_{2}$
\item[Case 2] $\kappa_{1}=\kappa_{2}$ and $\lambda_{1}\ne \lambda_{2}$
\item[Case 3] $\kappa_{1}=\kappa_{2}$, $\lambda_{1}= \lambda_{2}$, and $\mu_{1}\ne \mu_{2}$
\end{description}
In Case 1, we consider the case $\kappa_{1}<\kappa_{2}$. 
It follows from Lemma \ref{Proof:lemA1} that $\tilde{H}_{s}^{p}(\kappa)=O\left( p^{s}\right)$ $(s=1,3,5)$.
Hence, $S_{1,p}(\bm{\eta}_{1})^{2}$ increases in polynomial order of $p$. 

On the other hand, it follows from Lemma 9 of \citet{MSA19} that
\begin{align}
\frac{I_{p}(\kappa_{1})}{I_{p}(\kappa_{2})}=O\left( \left( \frac{\kappa_{1}}{\kappa_{2}}\right)^{p}\right) ,
\end{align}
which implies that $(I_{p}(\kappa_{1})/I_{p}(\kappa_{2}))^2$ decreases to zero in exponential order if $\kappa_{1}<\kappa_{2}$.
Therefore, we have
\begin{align}
\frac{\rho_{\textrm{MRL},p}(\bm{\eta}_{1})^{2}}{\rho_{\textrm{MRL},p}(\bm{\eta}_{2})^{2}}&=O(p^{2(2m+1)})O\left( \left( \frac{\kappa_{1}}{\kappa_{2}}\right)^{2p}\right) \to 0
\end{align}
as $p\to \infty$. The case with $\kappa_{2}<\kappa_{1}$ is also proved similarly. 

Second, we consider the case with $\kappa_{1}=\kappa_{2}=\kappa$ and $\lambda_{1}\ne \lambda_{2}$. We write $\alpha_{p}(\kappa )=\alpha_{p}$ and $\beta_{p}(\kappa ,\lambda)=\beta_{p}$ to clarify the dependence of parameters. Then, the characteristic function of the ESS-vM distribution is written by
\begin{align*}
\Psi (\bm{\eta})&=\exp (ip\mu )\left( \alpha_{p}(\kappa )+i\beta_{p}(\kappa ,\lambda )\right) .
\end{align*}

If $\lambda_{1}=0$, it is obvious that $\Psi (\bm{\eta}_{1})\ne \Psi (\bm{\eta}_{2})$ because $\beta_{p}(\kappa ,0)=0$. Therefore, without loss of generality, we assume $\lambda_{1}\ne 0$.
Now, we prove $\Psi (\bm{\eta}_{1})\ne \Psi (\bm{\eta}_{2})$ by contradiction.
To do that, we assume $\Psi (\bm{\eta}_{1})=\Psi (\bm{\eta}_{2})$.
It follows that
\begin{align}
\exp (ip\mu_{1} )\left( \alpha_{p}(\kappa )+i\beta_{p}(\kappa ,\lambda_{1} )\right) &=\exp (ip\mu_{2} )\left( \alpha_{p}(\kappa )+i\beta_{p}(\kappa ,\lambda_{2} )\right) ,\notag
\end{align}
where $-2\pi <\mu_{1}-\mu_{2}<2\pi$. 
%We consider the case $0\leq \mu_{1}-\mu_{2}<2\pi$. 
Then, 
\begin{align}
\exp (ip(\mu_{1}-\mu_{2}))&=\frac{ \alpha_{p}(\kappa )+i\beta_{p}(\kappa ,\lambda_{2} )}{\alpha_{p}(\kappa )+i\beta_{p}(\kappa ,\lambda_{1} )}\label{Proof:ESSvM_eq1} \\
 &=\frac{\alpha_{p}(\kappa )^{2}-\beta_{p}(\kappa ,\lambda_{2})\beta_{p}(\kappa ,\lambda_{1} )+i\left\{ \alpha_{p}(\kappa)\beta_{p}(\kappa ,\lambda_{2})-\alpha_{p}(\kappa)\beta_{p}(\kappa ,\lambda_{1})\right\}}{\rho_{\textrm{MRL},p}(\bm{\eta}_{1})^{2}}. \label{Proof:ESSvM_eq2}
\end{align}
The real part of expression \eqref{Proof:ESSvM_eq2} is
\begin{align}
&\frac{ \left\{ 1-S_{1,p}(\kappa ,\lambda_{1})S_{1,p}(\kappa ,\lambda_{2})\right\} I_{p}(\kappa)^{2}/I_{0}(\kappa)^{2}}{\left\{ 1+S_{1,p}(\kappa ,\lambda_{1})^2\right\}  I_{p}(\kappa)^{2}/I_{0}(\kappa)^{2}} \notag \\
&=\frac{ 1-S_{1,p}(\kappa ,\lambda_{1})S_{1,p}(\kappa ,\lambda_{2})}{1+S_{1,p}(\kappa ,\lambda_{1})^2}, \label{eqB.2}
\end{align}
where
\begin{align}
S_{1,p}(\kappa, \lambda)=C_{m}\sum_{\ell=0}^{m}\binom{m}{\ell}\frac{(-1)^{\ell +1}}{2\ell+1} \lambda^{2\ell +1} \sum_{k=0}^{\ell}c_{n(\ell ),n(\ell )-2k}\tilde{H}^{p}_{n(\ell)-2k}(\kappa).\notag
\end{align}
The order of $p$ in $\tilde{H}^{p}_{n(\ell)-2k}(\kappa)$ is maximal if and only if $\ell=m$ and $k=0$. Then, $\tilde{H}^{p}_{2m+1}(\kappa)=O(p^{2m+1})$. 
In addition, when $\ell=m$ and $k=0$, $c_{n(\ell),n(\ell)-2k}=c_{2m+1,2m+1}\ne 0$. Because for each $\ell$ and $k$ with $\ell =0,\ldots ,m-1$ and $k=0,\ldots ,\ell$, $\tilde{H}^{p}_{n(\ell )-2k}=o(p^{2m+1})$. 
Therefore, dividing the numerator and denominator in equation \eqref{eqB.2} by $(p^{2m+1})^2$ and taking the limit as $p\to\infty$ yield
\begin{align}
&\lim_{p\to\infty}\frac{ \left\{ 1-S_{1,p}(\kappa ,\lambda_{1})S_{1,p}(\kappa ,\lambda_{2})\right\}/(p^{2m+1})^2 }{\left\{1+S_{1,p}(\kappa ,\lambda_{1})^2\right\}/(p^{2m+1})^2}\notag \\
=&\lim_{p\to\infty}\frac{\left\{ -\frac{(-1)^{m+1}}{2m+1}\lambda_{1}^{2m+1}c_{2m+1,2m+1}\tilde{H}_{2m+1}^{p}(\kappa )/p^{2m+1}\right\}\left\{ -\frac{(-1)^{m+1}}{2m+1}\lambda_{2}^{2m+1}c_{2m+1,2m+1}\tilde{H}_{2m+1}^{p}(\kappa )/p^{2m+1}\right\}}{\left\{ -\frac{(-1)^{m+1}}{2m+1}\lambda_{1}^{2m+1}c_{2m+1,2m+1}\tilde{H}_{2m+1}^{p}(\kappa )/p^{2m+1}\right\}^2} \notag \\
=&\frac{\lambda_{2}^{2m+1}}{\lambda_{1}^{2m+1}}\ne 1. \notag
\end{align}
Next, we look at the left side of equation \eqref{Proof:ESSvM_eq1}, that is $\exp (ip(\mu_{1}-\mu_{2}))$. 
From \cite{MSA19}, there exists a subsequence $\{ p_{n}\}$ such that $p_{n}(\mu_{1}-\mu_{2})\to 0$ (mod $2\pi$) as $n\to\infty$.
Accordingly, we have 
\begin{align}
\exp (ip_{n}(\mu_{1}-\mu_{2})) \to 1 ,\qquad (n\to\infty ). \label{Proof:ESSvM_eq4}
\end{align}
As the limit $(\lambda_{2}/\lambda_{1})^{2m+1}$ also holds for the subsequence $\{ p_{n}\}$, this contradicts the result \eqref{Proof:ESSvM_eq4}.

Finally, we consider the case with $\kappa_{1}=\kappa_{2}=\kappa$, $\lambda_{1}= \lambda_{2}=\lambda$, and $\mu_{1}\ne \mu_{2}$. 
We prove $\Psi (\bm{\eta}_{1})\ne \Psi (\bm{\eta}_{2})$ by contradiction.
Assume $\Psi (\bm{\eta}_{1})=\Psi (\bm{\eta}_{2})$. Then, it follows that for any $p\in\mathbb{Z}$, 
\begin{align*}
\exp \left( ip(\mu_{1}-\mu_{2})\right)=1.
\end{align*}
Hence,
\begin{align*}
\frac{1}{N}\sum_{p=0}^{N-1}\exp \left( ip(\mu_{1}-\mu_{2})\right)=1.
\end{align*}
Taking the limit as $N\to\infty$ leads to $0=1$, which leads to contradiction.\hfill $\square$

\subsection{Proof of Theorem \ref{consistency_AN}}.
Assumption A4 implies that there exist some constants $M^* >0 $ and $m^*>0$ such that 
\begin{align*}
\sup_{\eta \in \bm{H}}  f_{\textrm{ESS}}^{(m)} (\theta; \bm{\eta}) < M^*
~~~\textrm{and}~~
\inf_{\eta \in \bm{H}}  f_{\textrm{ESS}}^{(m)}  (\theta; \bm{\eta}) > m^* ,
\end{align*}
which lead to $E_{\eta_0} [ \log f_{\textrm{ESS}}^{(m)} (\Theta; \bm{\eta})] > - \infty $. For every sufficiently small ball $U \subseteq \bm{H}$, 
\begin{align*}
E_{\eta_0} [  \sup_{\eta \in U}  \log f_{\textrm{ESS}}^{(m)} (\Theta; \bm{\eta})]< \infty.
\end{align*}
We write $\bm{H}_0$ for the set $\{ \bm{\eta}\in \bm{H} | E_{\bm{\eta}_0}[\log f_{\textrm{ESS}}^{(m)}(\Theta ;\bm{\eta}_0)] = \sup_{\eta}E_{\bm{\eta}_0}[\log f_{\textrm{ESS}}^{(m)}(\Theta; \bm{\eta})]\}$ of all points at which $E_{\bm{\eta}_0}[\log f_{\textrm{ESS}}^{(m)}(\Theta; \bm{\eta})]$ reaches its global maximum. From the Kullback-Leibler inequality and Assumption A7, it follows that $\bm{H}_0 =\{ \bm{\eta}_0\}$. Using Theorem 5.14 of \cite{VV20}, for estimator $\hat{\bm{\eta}}_{n}$ such that $\ell (\hat{\bm{\eta}}_{n}) \geq \ell (\bm{\eta}_0)$ and for every $\epsilon >0$, we have
\begin{align*}
P(||\hat{\bm{\eta}}_{n}-\bm{\eta}_0 || \geq \epsilon ) \to 0\qquad (n\to \infty),
\end{align*}
where $\| \cdot \|$ denotes Euclidean norm.
Next, we show asymptotic normality \eqref{AN}. To do so, we only prove that the absolute value of any third derivative of the log-likelihood $\ell_m(\bm{\eta})$ is bounded above by some constant. 
Without a loss of generality, we assume that $\bm{\rho}$ is a one-dimensional parameter. To simplify this exposition, the parameter vector is rewritten as $\bm{\eta}=(\eta_{1},\eta_{2},\eta_{3})^{T}$, where $\eta_{1}$, $\eta_{2}$, and $\eta_{3}$ correspond to $\mu$, $\rho$, and $\lambda$, respectively.
The third derivatives of the log-likelihood are defined as $\partial_{qrs}\ell_m(\bm{\eta}):=(\partial^{3}/\partial\eta_{q}\partial\eta_{r}\partial\eta_{s})\ell_m(\bm{\eta})$ $(q,r,s=1,2,3)$. 
When we let $S_m(x)=g_{m}(x)/G_{m}(x)$, the absolute values of the third derivatives $\partial_{qrs}\ell_m(\bm{\eta})$ are bounded above by a linear combination of 
\begin{align*}
Q_{\alpha,\beta}(\bm{\theta}_n,\bm{\eta}):=\frac{1}{n}\sum_{i=1}^{n}\frac{S_m(\lambda\sin(\theta_i -\mu))^{\alpha}}{\{ 1-\lambda^{2}\sin^{2}(\theta_i-\mu)\}^{\beta}},\quad (\alpha =0,1,2,3, \textrm{ and } \beta=0,1,2).
\end{align*}
Because the parameter space of $\lambda$ is $[-1+\delta_{\lambda}, 1-\delta_{\lambda}]$, we have
\begin{align*}
Q_{\alpha,\beta}(\bm{\theta}_n,\bm{\eta})\leq \frac{1}{1-(1-\delta_{\lambda})^2}\quad \textrm{ for any } \bm{\eta}\in\bm{H} \textrm{ and }\bm{\theta}_n .
\end{align*}
Hence,  $|\partial_{qrs}\ell_m(\bm{\eta})|\leq M^{\dag}$ for some positive constant $M^{\dag}$, and asymptotic normality holds from Theorem 5.41 of \cite{VV20}. Note that the constant $M^{\dag}$ corresponds to the fixed integrable function $\ddot{\psi}(x)$ in Theorem 5.41 of \cite{VV20}. \hfill $\square$

\subsection{Derivation of the random number generation method}
\begin{align}
P(\Theta \leq \xi)&=P\left(\Theta \leq \xi , U< G_{m}(\lambda\sin\Phi)\right) +P\left(\Theta \leq \xi , U\geq G_{m}(\lambda\sin\Phi)\right) \label{DR01}
\end{align}
Let $f(\phi ,u)$ be the joint density of $(\Phi ,U)$, and $f(u)$ be the density  of $U$. 
In the second term in the right hand side, it follows that
\begin{align}
P\left( -\Phi \leq \xi, U\geq G_{m}(\lambda\sin\Phi)\right) &=\int_{-\xi}^{\pi}\int_{G_{m}(\lambda\sin\phi)}^{1}f(\phi ,u)d\phi du \notag \\ 
&=\int_{-\xi}^{\pi}f_{0}(\phi )\left( \int_{G_{m}(\lambda\sin\phi)}^{1}f(u)du\right) d\phi \notag \\
&=\int_{-\xi}^{\pi}f_{0}(\phi )\left( 1-G_{m}(\lambda\sin\phi)\right) d\phi \label{DR02}
\end{align}
Because the density $g_{m}(\cdot )$ is symmetric about the origin, we have $1-G_{m}(\lambda\sin\phi)=G_{m}(-\lambda\sin\phi)$.
Using this result and the condition A1, Equation \eqref{DR02} becomes
\begin{align*}
\int_{-\xi}^{\pi}f_{0}(\phi )G_{m}(\lambda\sin (-\phi)) d\phi &=\int_{-\pi}^{\xi}f_{0}( -\phi^{\prime} )G_{m}(\lambda\sin \phi^{\prime}) (-1)d\phi^{\prime}\\
&=\int_{-\pi}^{\xi}f_{0}( \phi )G_{m}(\lambda\sin \phi)d\phi .
\end{align*}
Similarly, the first term in the right hand side in Equation \eqref{DR01} is written as $\int_{-\pi}^{\xi}f_{0}( \phi )G_{m}(\lambda\sin \phi)d\phi$. 
Therefore, we have $P(\Theta \leq \xi) =\int_{-\pi}^{\xi}2f_{0}( \phi )G_{m}(\lambda\sin \phi) d\pi$.

\end{document}